\documentclass[useAMS,usenatbib]{mn2e}
\newcommand{\feh}{\ensuremath{\left[{\rm Fe}/{\rm H}\right]}}
\newcommand{\teff}{\ensuremath{T_{\rm eff}}}

\newcommand{\msun}{\ensuremath{{\rm M}_\odot}}
\newcommand{\rsun}{\ensuremath{{\rm R}_\odot}}

\newcommand{\mj}{\ensuremath{{\rm M}_{\rm J}}}
\newcommand{\rj}{\ensuremath{{\rm R}_{\rm J}}}

\usepackage[pdftex]{graphicx}
\usepackage{amsmath}
\usepackage{amssymb}
\usepackage{deluxetable}

\title[Evidence for aerosols in the atmosphere of WASP-12b]{HST hot Jupiter
  transmission spectral survey:  
evidence for aerosols and lack of TiO in the atmosphere of WASP-12b}

\author[D. K. Sing et al.]
{D. K. Sing$^{1}$\thanks{E-mail: sing@astro.ex.ac.uk},  
A. Lecavelier des Etangs$^{2}$, 
J. J. Fortney$^{3}$,
A. S. Burrows$^{4}$, 
F. Pont$^{1}$, \newauthor
H. R. Wakeford$^{1}$,
G. E. Ballester$^{5}$, 
N. Nikolov$^{1}$, 
G. W. Henry$^{6}$,
S. Aigrain$^{7}$,
D. Deming$^{8}$, \newauthor
T. M. Evans$^{7}$,
N. P. Gibson$^{9}$,
C. M. Huitson$^{1}$,
H. Knutson$^{10}$,
A. P. Showman$^{7}$,\newauthor
A. Vidal-Madjar$^{2}$,
P. A. Wilson$^{1}$,  
M. H. Williamson$^{5}$,
K. Zahnle$^{11}$
\\
$^{1}$Astrophysics Group, School of Physics, University of Exeter, Stocker
  Road, Exeter, EX4 4QL\\ 
$^{2}$CNRS, Institut dAstrophysique de Paris, UMR 7095, 98bis boulevard Arago, 75014 Paris, France\\
$^{3}$Department of Astronomy and Astrophysics, University of
California, Santa Cruz, CA 95064, USA\\
$^{4}$Department of Astrophysical Sciences, Peyton Hall, Princeton University, Princeton, NJ 08544, USA\\ 
$^{5}$Lunar and Planetary Laboratory, University of Arizona, Tucson, Arizona 85721, USA\\
$^{6}$Tennessee State University, 3500 John A. Merritt Blvd., P.O. Box 9501, Nashville, TN 37209, USA\\ 
$^{7}$Department of Physics, University of Oxford, Denys Wilkinson Building, Keble Road, Oxford OX1 3RH, UK\\ 
$^{8}$Department of Astronomy, University of Maryland, College Park, MD 20742 USA\\
$^{9}$European Southern Observatory, Karl-Schwarzschild-Str. 2, 85748 Garching bei Munchen, Germany\\ 
$^{10}$Division of Geological and Planetary Sciences, California Institute of Technology, Pasadena, CA 91125 USA\\ 
$^{11}$NASA Ames Research Center, Moffett Field, CA 94035, USA
}
\begin{document}

\date{Accepted 2013 September 19.  Received 2013 September 4; in original form 2013 July 29}

\pagerange{\pageref{firstpage}--\pageref{lastpage}} \pubyear{2013}
\maketitle

\label{firstpage}

\begin{abstract}
We present {\it Hubble Space Telescope} ({\it HST}) optical transmission spectra
of the transiting hot Jupiter WASP-12b, taken with the Space Telescope
Imaging Spectrograph (STIS) instrument.  The resulting spectra cover
the range 2900 to 10300 \AA\, which we combined with archival WFC3
spectra and {\it Spitzer} photometry to cover the full optical to infrared
wavelength regions.   
With high spatial resolution, we are able to resolve WASP-12A's stellar companion in both our images and spectra, revealing
that the companion is in fact a close binary M0V pair, with the three stars forming a triple-star configuration.
We derive refined physical parameters of the WASP-12 system, including
the orbital ephemeris, finding the exoplanet's density is $\sim$20\% lower than previously
estimated.  

From the transmission spectra, we are able to decisively rule out prominent absorption by 
TiO in the exoplanet's atmosphere, as there are no signs of the
molecule's characteristic broad features nor
individual bandheads.  Strong pressure-broadened Na and K
  absorption signatures are also excluded, as are significant
  metal-hydride features.
We compare our combined broadband spectrum to a wide variety of existing
aerosol-free atmospheric models, though none are satisfactory fits. 
However, we do find that the full transmission spectrum can be
described by models which include significant opacity
from aerosols: including Rayleigh scattering,
Mie scattering, tholin haze, and settling dust profiles.
The transmission spectrum follows an effective extinction cross section
with a power-law of index $\alpha$, with
the slope of the transmission spectrum constraining the quantity $\alpha T = -3528\pm$660 K,
where $T$ is the atmospheric temperature. 
Rayleigh scattering ($\alpha=-4$) is among the best fitting models, though requires
low terminator temperatures near 900 K.  
Sub-micron size aerosol particles can provide equally good fits to the entire transmission spectrum
for a wide range of temperatures, and we explore corundum as a plausible dust aerosol.
The presence of atmospheric aerosols also helps to explain the
modestly bright albedo
implied by {\it Spitzer} observations, as
well as the near black body nature of the emission spectrum.
Ti-bearing condensates on the cooler night-side is the most 
natural explanation for the overall lack of TiO signatures in WASP-12b,
indicating the day/night cold-trap is an important effect for very hot Jupiters.
These finding indicate that aerosols can play a significant atmospheric
role for the entire wide range of hot-Jupiter
atmospheres, potentially affecting their overall spectrum and energy balance. 

\end{abstract}


\section{Introduction}
Transit events have revolutionised our understanding of hot-Jupiter
exoplanets, thanks in large part to atmospheric studies using
transmission, emission, and phase curve observations.
Even with hundreds of known transiting hot Jupiters, currently only a
small handful of exoplanets are capable of having their atmosphere characterised with
all three types observations using current facilities, making these
planets prototypes for the rest of the field.  Among these planets is
WASP-12b, a prototype `very hot Jupiter ' which is one of the
largest and hottest planets known \citep{2009ApJ...693.1920H}. 
Extensive ground and space-based observations have been conducted on
this planet over the last few years
\citep{2010ApJ...716L..36L, 
2010ApJ...714L.222F, 
2011AJ....141..179C, 
2011ApJ...727..125C, 
2011AJ....141...30C, 
2011AA...528A..65M, 
2012ApJ...747...82C,  
2012ApJ...760...79H, 
2013Icar..225..432S, 
2012ApJ...760..140C, 
2012ApJ...746...46C, 
2012ApJ...748L...8Z, 
2013ApJ...766L..20F, 
2013A&A...551A.108M, 
2013MNRAS.434..661C, 
2013arXiv1305.1670S, 
2013arXiv1308.0337F}.  

As a very hot planet, most models predict strong Na, K, H$_2$O, and TiO/VO features
in the atmosphere, with no condensates.
WASP-12b has the potential to help address the stratospheric TiO
hypothesis, 
which links the presence of TiO to thermal inversions \citep{2008ApJ...678.1419F}. As one of the largest
planets, studying its atmosphere could also help gain insight into the connection
between the anomalously large radius and atmospheric features. 

Despite all the recent observations, the nature and composition of WASP-12b's
lower atmosphere remains largely elusive, as both ground and space-based
transmission spectra have not yet been able to place
strong constraints on the presence of specific molecular features or
atmospheric constituents
\citep{2013Icar..225..432S, 2013MNRAS.434..661C, 2013arXiv1305.1670S}. 
Spectral retrieval methods on the day-side emission
spectrum indicated the planet contains a high C/O ratio \citep{2011Natur.469...64M}, which would have important
consequences regarding the planet's atmosphere
\citep{2012ApJ...758...36M}.  
However, WASP-12A is now known to have a close M-dwarf stellar
companion discovered by \cite{2011IAUS..276..397B, 2013MNRAS.428..182B}, which
diluted previously reported transit and eclipse depths, especially in the infrared.  
After the diluting effects of the nearby M-dwarf have been taken into
account, the dayside spectrum appears to largely resemble a black body 
\citep{2012ApJ...760..140C, 2013Icar..225..432S}, 
which weakens the case for a high C/O ratio atmosphere.

Here we present results from a large {\it Hubble Space Telescope}
({\it HST}) programme
for WASP-12b ({\it HST} GO-12473; P.I. Sing), which is part of an eight planet optical atmospheric survey of transiting
hot Jupiters, with the initial results given in
\cite{2013arXiv1307.2083H} 
for WASP-19b and \cite{2013arXiv1308.2106W, Nikolov2013} for HAT-P-1b.
The overall programme goals are to detect atmospheric features across a
wide range of hot-Jupiter atmospheres enabling comparative
exoplanetology, detect stratosphere causing agents like TiO
\citep{2003ApJ...594.1011H, 2008ApJ...678.1419F}, 
and detect hazes and clouds.
In this paper, we present new {\it HST} transit observations with the STIS
instrument (Space Telescope Imaging Spectrograph), and combine them
with existing WFC3 (Wide Field Camera 3) spectra and {\it Spitzer}
photometry to construct a high signal-to-noise near-UV to infrared
transmission spectrum, capable of detecting and scrutinising atmospheric constituents.
We describe our observations in Sect. 2, present the analysis of the
transit light curves in Sect. 3, discuss the results in Sect. 4 and
conclude in Sect. 5.


\section{Observations}
\subsection{\emph{Hubble Space Telescope} STIS spectroscopy}
We observed two transits of WASP-12b with the {\it HST} STIS G430L
grating during 14 March 2012 and 27 March 2012, as well as one transit
with the STIS G750L during 4 September 2012.  
The G430L and G750L datasets 
all consist of 53 spectra, each spanning five spacecraft orbits.
 The G430L grating covers
the wavelength range from 2,900 to 5,700~{\AA}, with a resolution R of
$\lambda$/$\Delta\lambda=$530--1,040 ($\sim$2 pixels; 5.5~{\AA}).  
The G750L grating covers
the wavelength range from 5,240 to 10,270 ~{\AA}, with a resolution R of
$\lambda$/$\Delta\lambda=$530--1,040 ($\sim$2 pixels; 9.8~{\AA}).  
Both the G430L and G750L data were taken with a wide 52$''\times2''$ slit to minimise slit light
losses.  
This observing technique has proven to
produce high signal-to-noise (S/N) spectra which are photometrically accurate near the 
Poisson limit during a transit event (e.g
\citealt{2001ApJ...553.1006B, 2011MNRAS.416.1443S, 2012MNRAS.422.2477H}).  
The three visits of {\it HST} were scheduled such that the third and fourth spacecraft orbits
would contain the transit, providing good
coverage between second and third contact, as well as an out-of-transit
baseline time series before and after the transit. 
Exposure times of 279 seconds were used in conjunction with a 128-pixel wide
sub-array, which reduces the readout time between exposures
to 21 seconds, providing a 93\% overall duty cycle.

The dataset was pipeline-reduced with the latest version of {\tt CALSTIS} 
and cleaned for cosmic ray detections with custom {\tt IDL} routines.
Cosmic rays typically affect $\sim 5\%$ of the CCD pixels in a
typical 279 second exposure and the STScI pipeline is generally insufficient
in cleaning the images.  Therefore, we adopted the strategy developed by
\cite{Nikolov2013} to clean the images.  The procedure is based on median-combining difference images to
identify cosmic ray events, substituting the affected pixels, and those
identified by {\tt CALSTIS} as ``bad'', with flux values interpolated from
the nearby point spread function (PSF).
The mid-time of each exposure was converted into BJD$_{TBD}$ for use
in the transit light curves \citep{2010PASP..122..935E}. 
As the G750L grating suffers from fringing effects red-ward of
$\sim$7250\AA, we obtained contemporaneous fringe flats at the end of
observing sequence, using them to defringe the science frame images
(also see \citealt{Nikolov2013}).

\begin{figure*}
 {\centering
  \includegraphics[width=1\textwidth,angle=0]{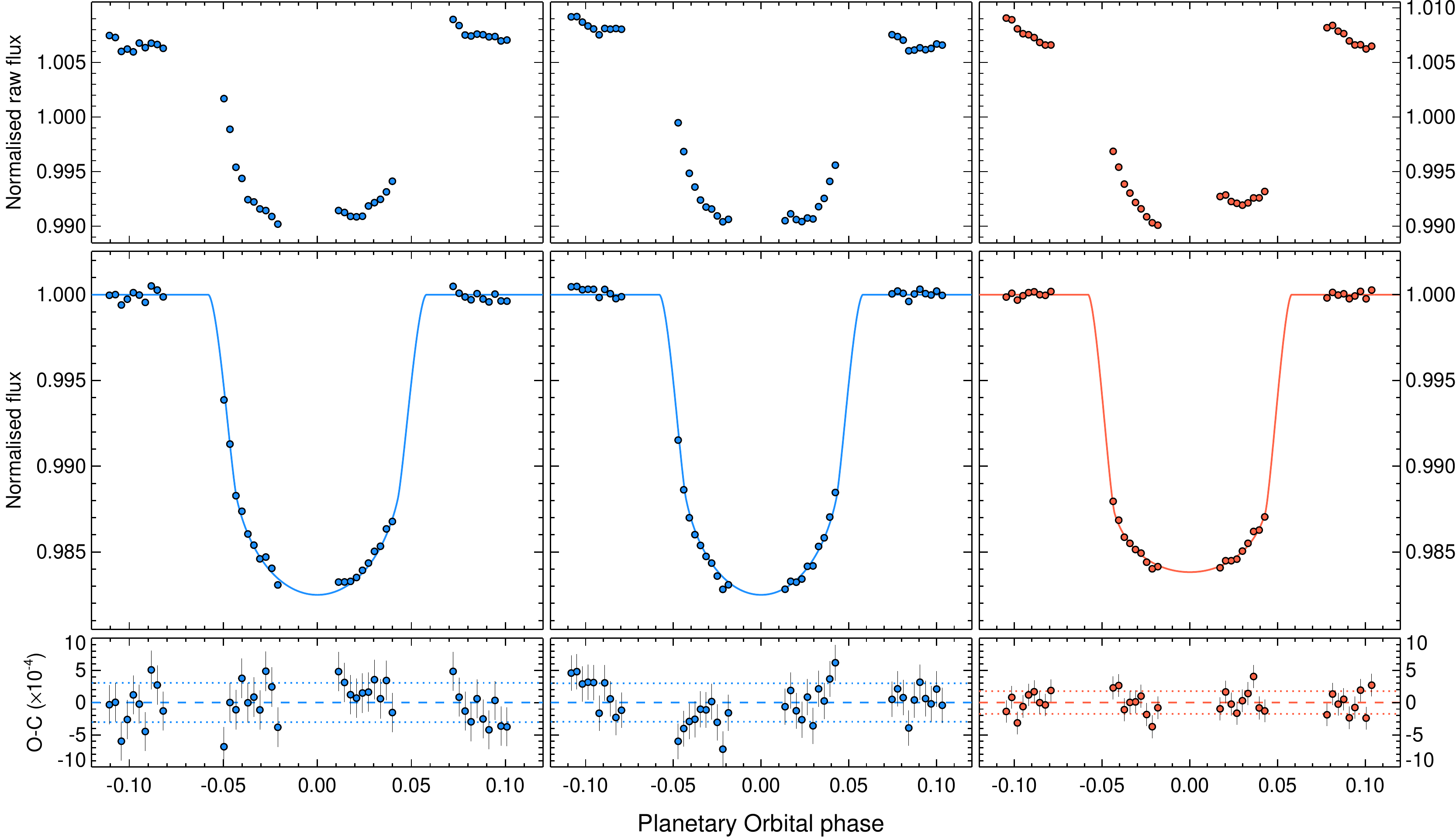}}
\caption[]{{\it HST} STIS normalised white-light curves for the three
  WASP-12b transits taken with the G430L (blue - left \& middle) and
  G750L (red - right). 
Top panels: Raw light curves normalised to the
  mean raw flux. The light curves experience
  prominent systematics associated primarily with the {\it HST}
  thermal-breathing cycle (see text for details). 
Middle panels: Detrended light curves along with the best-fit
limb-darkened transit model superimposed with continuous lines.
Lower panels: Observed minus modelled light curve residuals, compared
to a null (dashed lines) and 1$-\sigma$ level (dotted lines).
A colour version is available in the online version of the journal.}
\label{figwhite}
\end{figure*}

The spectral aperture extraction was done with {\tt IRAF} using a
13-pixel-wide aperture with no background subtraction, which
  minimises the out of transit standard deviation of the white light curves.
The extracted spectra were then Doppler-corrected to a common rest frame through cross-correlation, 
which removed sub-pixel wavelength shifts in the dispersion direction. 
The STIS spectra were then used to create both a white-light photometric 
time series (see Fig. \ref{figwhite}), and custom wavelength bands covering the spectra, 
integrating the appropriate wavelength flux from each exposure for
different bandpasses.  The resulting photometric light curves
exhibit all the expected systematic instrumental effects taken during similar
high S/N transit observations before {\it HST} Servicing Mission 4 with STIS, as first noted
in \cite{2001ApJ...553.1006B}. 

The main instrument-related systematic effect is primarily due to the well known
thermal breathing of {\it HST}, which warms and cools the telescope during the 96 minute 
day/night orbital cycle, causing the focus to vary\footnote{see STScI
  Instrument Science Report ACS 2008-03}.  Previous observations have
shown that once the telescope is slewed to a new pointing position,
it takes approximately one spacecraft orbit to thermally relax, which
compromises the photometric stability of the first orbit of each {\it HST} visit.  In
addition for STIS, the first exposure of each spacecraft orbit has consistently
been found to be significantly fainter than the remaining exposures.  These trends are continued
in our STIS observations, and are both minimised in the analysis with
proper {\it HST} visit scheduling.  Similar to other studies, in our
subsequent analysis of the transit we discarded the first orbit
of each visit (purposely scheduled well before the transit event).  In addition, we set the
exposure time of the first exposure of each spacecraft orbit to be 1
second in duration, such that the exposure could be discarded without
significant loss in observing time.  We find that the second exposures taken
for all five spacecraft orbits (each 279 seconds in duration) do not show
the first-exposure systematic trend, making the short-exposure strategy
an effective choice.  However, we note that the first-exposure
systematic in new observations of HD~189733 with the STIS G430L (GO
1306, P.I. Pont), persisted despite discarding similar 1 second initial exposures.  As the
HD~189733 time series exposures were much shorter (64 seconds), the
1-second strategy may only be effective for fainter targets like
WASP-12, which have longer exposure times.

\subsection{\emph{Hubble Space Telescope} WFC3 spectroscopy}
We also analyse the WFC3 G141 transit data of WASP-12 taken as part of GO 12230
(PI Swain), with results given in
\cite{2013Icar..225..432S}.  
Our re-analysis is motivated by both a need
for a consistently derived transmission spectrum with the STIS data,
and the subsequent discovery of the M-dwarf companion
\citep{2011IAUS..276..397B, 2013MNRAS.428..182B}
which affects the transit and eclipse depths
\citep{2012ApJ...760..140C}. 

We use the {\tt flt} outputs from WFC3's {\tt calwf3} pipeline.  For each
exposure {\tt calwf3} conducts the following processes: bad pixel
flagging, reference pixel subtraction, zero-read subtraction, dark
current subtraction, non-linearity correction, flat-field correction,
as well as gain and photometric calibration. The resultant images are
in units of electrons per second.   The first orbit of the WFC3 data
was removed, as it also suffers from thermal breathing systematic
effects like STIS, leaving 391 exposures over the remaining four {\it HST}
orbits, with a total of 175 exposures taken between first and fourth contact. 

The spectra were extracted using custom {\tt IDL} routines similar to
{\tt IRAF}'s {\tt APALL}
procedure, using an aperture of $\pm$9 pixels from the central row, 
determined by minimising the standard deviation across the
aperture.  
The resulting 18 pixel wide aperture encompasses both WASP-12 and the M-dwarf 
binary.  The aperture was traced around a computed centering profile, found to be 
consistent in the y-axis with an error of $\pm$0.5
pixels.  Background subtraction was applied using a region 
above and below the spectrum for each exposure. 
We elected to subtract the M-dwarf flux contribution 
through post-processing, though we note that nearly identical
results can be obtained by instead excluding the M-dwarf flux using PSF
fitting techniques \citep{2013Icar..225..432S}. 

For wavelength calibration, direct images were taken in the F132N
narrow band filter at the beginning of the observations providing the
absolute position of the target star.   We
assumed that all pixels in the same column have the same 
effective wavelength, with the extracted spectra covering the
wavelength range of 1.077 to 1.704~$\mu$m.

\subsection{Photometric Monitoring for
Stellar Activity}
\label{section:activity}
\begin{figure}
 {\centering
  \includegraphics[width=0.47\textwidth,angle=0]{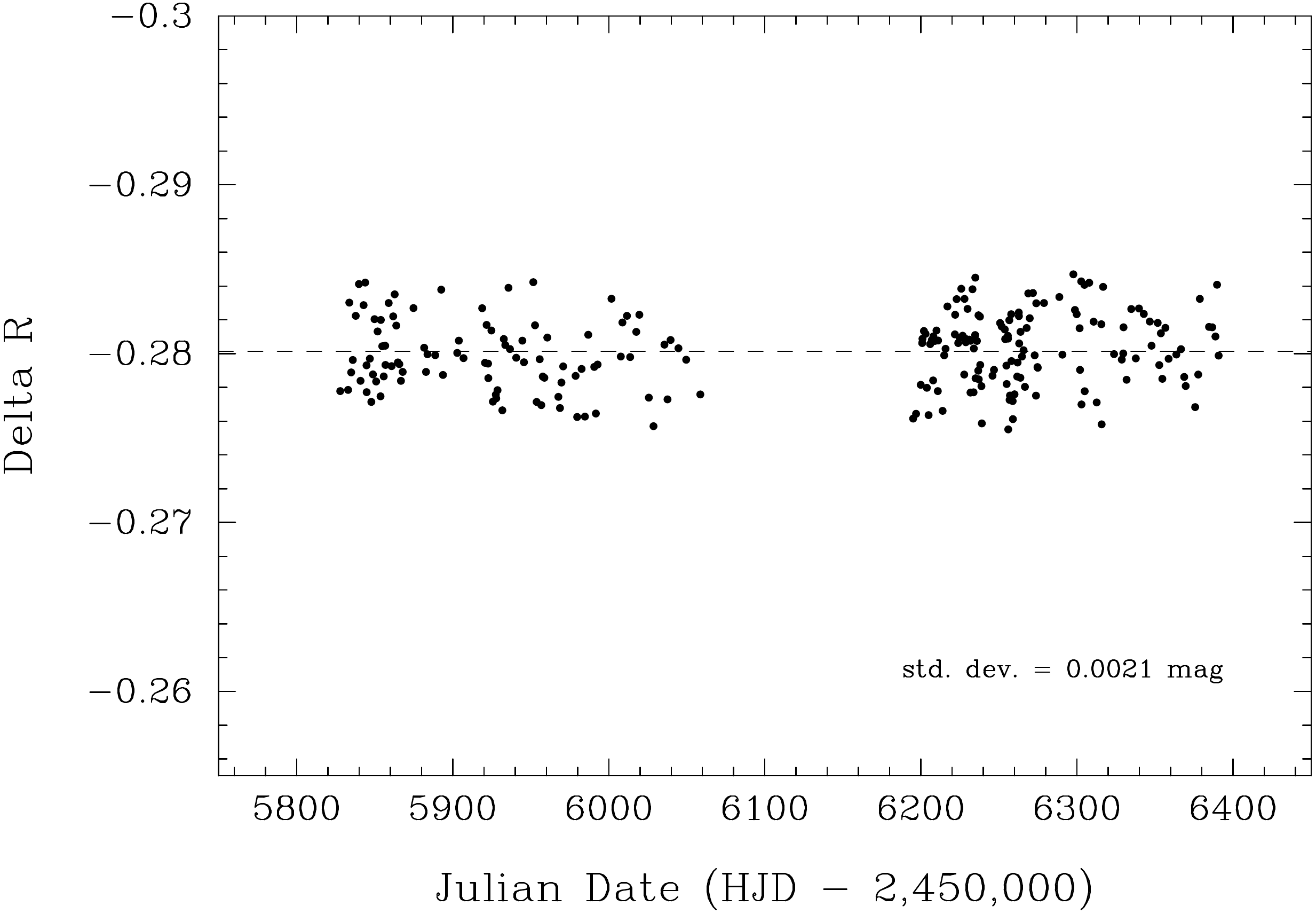}}
\caption[]{Stellar activity monitoring of WASP-12A from the Fairborn
  Observatory, showing the differential R magnitude from 2011 to 2013
  with transits removed.  The standard deviation of the light curve is 2.1
  mmag, with no significant activity nor periodicity detected.}
\label{Figure:Monitor}
\end{figure}

We also observed WASP-12 from the ground with Tennessee State University's
automated Celestron C-14 telescope, located at Fairborn Observatory in
southern Arizona.
The telescope was equipped with an SBIG STL-1001E CCD camera observing
through a Cousins R filter.
Fig. \ref{Figure:Monitor} shows the Cousins R data for the 2011-2012 and 2012-2013
observing seasons plotted against Heliocentric Julian Date.  The observations have been corrected for bias, flat field,
differential extinction, and pier-side offset.  
Differential magnitudes are computed against the mean of 17 comparison
stars in the same CCD field.  Each differential magnitude plotted is
the mean of 8 to 10 successive frames on a given night.  There are a
total of 231 nightly observations, with transit points excluded from
the plot. The standard deviation
of an individual observation from the grand mean is 2.1 mmag.  We
find no significant periodicity between one hour and 200 days in these
data, with WASP-12A appearing quite constant.
In particular, a least-squares sine fit of the C14 observations phased
with the radial velocity period has a peak-to-peak amplitude of only
$0.00056~\pm~0.00041$ mag, indicating little, if any, contribution to
the radial velocity variations from stellar-surface activity.
The late F spectral type of WASP-12 and its relatively low measured $v$sin$i$ \citep{2010ApJ...720.1569K},
also makes it unlikely to have significant star spots, which is confirmed by our photometric monitoring.
Alternatively, the low $v$sin$i$ could also be a result of a near
  pole-on view of the star as seen from the Earth, which is consistent
  with the spin-orbit misalignment measurement from \cite{2012ApJ...757...18A}.  

\section{Analysis}
\label{Sec:analysis}

\subsubsection{STIS white-light curve fits}
\label{Sec:STISwhite}
 Our overall analysis methods are similar to the {\it HST} analysis of
\cite{2011MNRAS.416.1443S, 2013arXiv1307.2083H} 
and \cite{Nikolov2013}, which we also describe briefly here. 
The light curves were modelled with the analytical transit
models of \cite{
2002ApJ...580L.171M}.   
For the white-light curves, the 
central transit time, inclination, stellar density, planet-to-star radius
contrast, stellar baseline flux, and instrument systematic trends were
fit simultaneously, with flat priors assumed.  The period was initially fixed to a literature value, before
being updated, with our final fits adopting the value from section \ref{Sec:ephemeris}.  Both G430L transits were fit with a common
inclination, stellar density and planet-to-star radius contrast.  The
results from the G430L and G750L white light curve fits were then used in conjunction with literature
results to refine the orbital ephemeris and
overall planetary system properties (see section \ref{Sec:jointfit}).  
To account for the effects of limb-darkening on the transit light
curve, we adopted the four parameter non-linear limb-darkening law, 
calculating the coefficients with ATLAS stellar models ($T_{eff}$=6500, log~$g$=4.5, [Fe/H]= 0.0) following \cite{
2010A&A...510A..21S}. 
Varying the stellar model within the known parameter range has a small effect on
  the output limb-darkening coefficients and fit transit parameters.  In particular, we find changing the stellar effective
temperature by $\pm$250~K shifts the radius ratios by
$\sim$1/3-$\sigma$.  Further, the shift is typically common to all
wavelength channels, and the relative radius ratio differences are largely preserved
(to levels of $\sim$0.1-$\sigma$ in the case of WASP-12b).

As in many past STIS studies, we applied orbit-to-orbit flux corrections by
fitting for a fourth-order polynomial to the photometric time series,
phased on the {\it HST} orbital 
period.  The systematic trends were fit simultaneously with the
transit parameters in the fit.  Higher-order polynomial fits were not statistically justified, based upon the
Bayesian information criteria (BIC; \citealt{Schwarz1978}).  The baseline flux level of each
visit was let free to vary in time linearly, described by two fit
parameters.  In addition, similar to the STIS analysis of \cite{2011MNRAS.416.1443S, 2012MNRAS.422.2477H} and \cite{2013arXiv1307.2083H}, 
we found it justified by the BIC to also linearly fit for
two further systematic trends which correlated with the X and Y detector positions of the spectra,
as determined from a linear spectral trace in {\tt IRAF}, as well as the wavelength shift between spectral
exposures, measured by cross-correlation.

The errors on each datapoint were initially set to the pipeline values which
is dominated by photon noise but also includes readout noise.
The best-fit parameters were determined 
simultaneously with a Levenberg-Marquardt least-squares algorithm 
\citep{2009ASPC..411..251M} 
using the unbinned data.
After the initial fits, the uncertainties for each data point were
rescaled based on the standard deviation of the residuals
and any measured systematic errors correlated in time (`red noise'), thus taking into account any underestimated
errors in the datapoints.  
The red noise was measured by checking whether the binned residuals followed a $N^{-1/2}$ relation, when
binning in time by $N$ points.  In the presence of red noise, the
variance can be modelled to follow a $\sigma^2=\sigma_w^2/N+\sigma_r^2$ relation,
where $\sigma_w$ is the uncorrelated white noise component, and
$\sigma_r$ characterises the red noise \citep{
2006MNRAS.373..231P}.  
We find that the pipeline per-exposure errors are accurate
at small wavelength bin sizes, which are dominated by photon noise, but are in general an underestimation
at larger bin sizes.  For the STIS white light curves, we find $\sigma_r=0.00008$ for
  the G430L and $\sigma_r=0$ for the G750L. 
A few deviant points from each light curve were
cut at the 3-$\sigma$ level in the residuals.


\subsubsection{WFC3 white-light curve fits}
\label{Sec:wfc3white}
\begin{figure}
 {\centering
  \includegraphics[width=0.45\textwidth,angle=0]{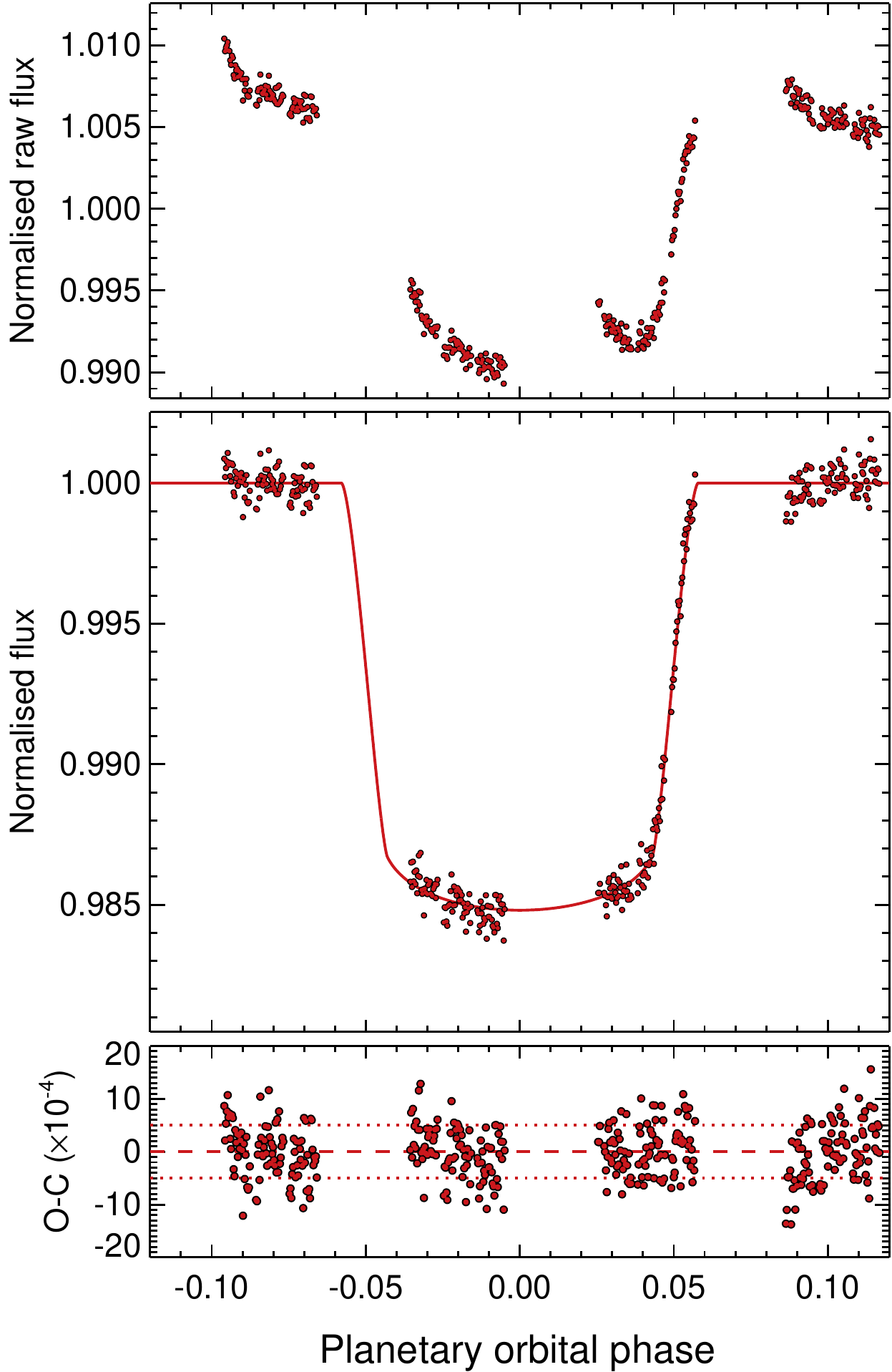}}
\caption[]{Similar to Fig. \ref{figwhite} but for the {\it HST} WFC3 transit.}
\label{figwhitewfc3}
\end{figure}

The noise properties and systematics of WFC3 G141 spectra are becoming
fairly well understood, and are mainly affected by  the
{\it HST} thermal-breathing variation (note the similarities of Figs. \ref{figwhite}
\& \ref{figwhitewfc3}) and detector persistence which
causes a ramp-effect, also known as the `hook'
\citep{2012ApJ...747...35B, 
2013ApJ...774...95D, 
2013Icar..225..432S}.  
These trends are predominantly either
common-mode (i.e. the same in every wavelength channel across the detector) or repeatable orbit-to-orbit, and thus straightforward to
remove in a variety of ways without affecting the relative transit
depths \citep{2012ApJ...747...35B, 
2013ApJ...774...95D, 
2013arXiv1307.2083H}.  
The amplitude of the `hook' has been studied as a
function of exposure level, and reported in \cite{2013ApJ...774...95D}.  
The authors found that the effect is practically zero when the exposure level per
frame is lower than 30,000 electrons/pixel, which is also well within
the linear regime of the detector.  As the maximum count rates for the
WASP-12 WFC3 data are in the range of 38,000 electrons/pixel, thus not
substantially higher than the nominal lower threshold, no
obvious `hook' systematics are observed (see also \citealt{2013Icar..225..432S}).

We modelled the WFC3 white-light curve using the same analytic transit
model and
limb-darkening stellar model as used in the STIS analysis.  Before
the model fits, we subtracted the flux contribution from the M-dwarfs
calculated from the WFC3 G141 response and the flux ratio as determined by \cite{2012ApJ...760..140C}, 
which matches well with our STIS analysis of both the M-dwarf spectral
type and flux normalisation (see Sec. \ref{Sec:Mdwarf}).  
We excluded the ends of the spectrum, where the response drops rapidly,
extracting the region from 1.137 to 1.657 $\mu$m to use for the
white-light curve.  

Like in the STIS analysis, the
systematic effects of the thermal-breathing were modelled with a fourth
order polynomial, phased on the {\it HST} orbital 
period, with the stellar baseline flux free to vary linearly in
time.  Unlike \cite{2013Icar..225..432S} 
we find that higher orders in {\it HST} orbital phase are required to more fully
remove the systematic errors.
An alternative method has been developed by
\cite{2012ApJ...747...35B} 
to correct for detector systematics, which takes
advantage of the repetitive nature.  Dubbed {\tt divide-OOT}, the procedure
involves creating a template baseline-flux time-series for each
subsequent exposure of an {\it HST}
orbit, determined by averaging the out-of-transit exposures (OOT).  The
light curve for each of the {\it HST} orbits are then divided by the OOT
time-series, which removes the systematic error components which are
common to each orbit.  We find {\tt divide-OOT} gives very similar results as the
parameterised method (as WFC3 exhibits largely common-mode systematics),
though shows somewhat higher levels of correlated noise,
as there are {\it HST} orbital phase mismatches between the
sub-exposures of the five {\it HST} orbits, leading to residual thermal breathing
trends which are evident in the residuals.  We adopt the
parameterised method both because of the somewhat lower noise values
($\sigma_r=0.00004)$,
and it is more straightforward to budget for the affects of
the dominant thermal-breathing systematic
on the transit depths through the use of the covariance matrix (or
alternatively the MCMC posterior distribution).
The M-dwarf flux contribution in the WFC3 is
$7.16\pm0.31\%$, which translates to a systematic uncertainty on the radius
ratio when subtracting the flux of 0.00018 $R_{p}/R_{*}$, which is small compared
to our final WFC3 white-light curve radius and uncertainly of 0.11977$\pm$0.00042 $R_{p}/R_{*}$.

In comparing our results to the literature, a source of disagreement is the overall $R_{p}/R_{*}$ level,
which we find significantly higher than \cite{2013arXiv1305.1670S}, 
with our levels much closer to those of \cite{2013Icar..225..432S}. 
As noted by \cite{2013arXiv1305.1670S},  
this is a result of the exponential function with time which they used to model the instrument
systematics, which effectively `carves out' the transit baseline flux
leading to smaller fit radii.  We see no evidence for a time-dependent exponential function (see
Fig. \ref{figwhitewfc3}), with our raw WFC3 white light curve
showing a closer resemblance to the STIS G750L data and its thermal-breathing trends.

\subsection{Transit ephemeris}
\label{Sec:ephemeris}

We used the central transit times of our {\it HST} data along with the
transit times of 
\cite{2009ApJ...693.1920H}, 
\cite{2011AJ....141..179C}, 
\cite{2011AA...528A..65M},  
and \cite{2012ApJ...747...82C} 
to determine an updated transit ephemeris.  
When calculating transit ephemerides from data taken with different
instruments in different wavelengths, it is important to use coherent
and realistic estimates.  Pont et al. (2006) showed how correlated
noise could dominate the error budget of system parameter measurements
from transit photometry.  
This issue is especially important for ground-based photometry of WASP-12, given that the
M0 dwarf companions lie 1$''$ from the target star, making very high
precision photometry more difficult in changing seeing conditions.
As the uncertainties given by
\cite{2011AA...528A..65M} 
do not account for the effect of correlated noise,
as a conservative estimate, we multiplied the
error bars by a factor of three, found to be typical for ground-based photometry of transiting planets (see also \citealt{2013MNRAS.430.3032B}).  

\begin{figure}
 {\centering
  \includegraphics[width=0.47\textwidth,angle=0]{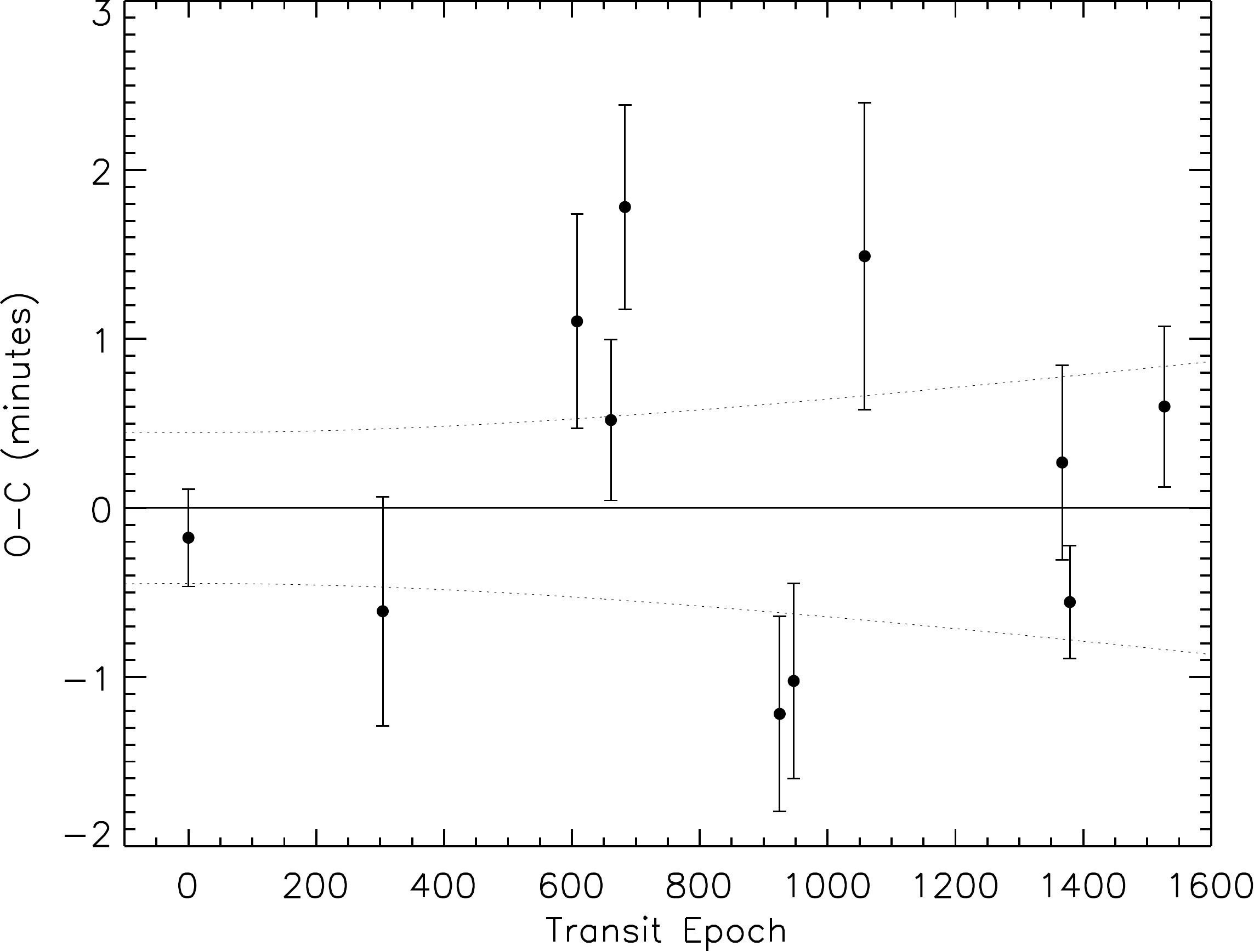}}
\caption[]{O-C diagram for the transit times of WASP-12 using the
  transits listed in Table \ref{Table:ttimes} and a linear ephemeris.
The 1$\sigma$ error envelope on the ephemeris is plotted as the dotted lines.}
\label{Figure:Timing}
\end{figure}

We fit the transit times listed in Table \ref{Table:ttimes} and shown
in Fig. \ref{Figure:Timing} using a linear function of the
period $P$ and transit epoch $E$,
\begin{equation} T(E) = T(0) + EP. \end{equation}
We find a
period of $P = 1.09142113\pm0.00000032$ (days)
and a mid-transit time of $T(0)=2454508.977005\pm$0.00031 (BJD).
A fit with a linear ephemeris is in general an unsatisfactory fit to the data,
having a $\chi^2$ of 29 for 9 degrees of freedom ($\chi^2_{\nu}$=3.2), and
transit timing variations (TTVs) have been claimed for this system
\citep{2013A&A...551A.108M}.  
We tentatively attribute the high formal reduced $\chi^2$ to
unaccounted red noise in the ground-based photometry, rather than
actual transit timing variations.
We proceed with the linear ephemeris given that our {\it HST} transit
times are more consistent with a linear fit than a TTV,
and the overall problems concerning the red-noise in photometry.  

\begin{table} 
\caption{Transit Timing for WASP-12b}
\label{Table:ttimes}
\begin{centering}
\renewcommand{\footnoterule}{}  
\begin{tabular}{llll}
\hline\hline  
 Epoc & BJD$_{TBD}$      &                    BJD error & Notes\\
\hline\hline  
    0   & 2454508.97688  & 0.00020    &\cite{2009ApJ...693.1920H}\\  
 304  & 2454840.76860  & 0.00047    &\cite{2011AJ....141..179C}\\ 
 608  & 2455172.56182  & 0.00044    &\cite{2011AJ....141..179C}\\ 
 661  & 2455230.40673  & 0.00033    &\cite{2011AA...528A..65M}\\ 
 683  & 2455254.41887  & 0.00042     &\cite{2011AA...528A..65M}\\ 
 925  & 2455518.5407    & 0.0004       &\cite{2012ApJ...747...82C}\\ 
 947  & 2455542.5521    & 0.0004       &\cite{2012ApJ...747...82C}\\
1058 & 2455663.70159  & 0.00063     &{\it HST} WFC3\\
1367 & 2456000.94987  & 0.00040     &{\it HST} G430L\\
1379 & 2456014.04635  & 0.00023     &{\it HST} G430L\\
1527 & 2456175.57748  & 0.00033     &{\it HST} G750L\\
\hline
\end{tabular}
\end{centering}

\end{table}

\subsection{System parameters}
\label{Sec:jointfit}

The M-dwarf stellar companion is fully resolved with respect to the WASP-12
primary in our STIS spectroscopy.  In combination with the high
signal-to-noise, this permits a precise determination of the planetary
radius completely free of the flux-dilution effects of the M-dwarf pair.  Given
that WASP-12b still has one of the largest radii of any transiting
planet, a precise radius for this exoplanet, in particular, is crucial
for understanding the hot-Jupiter inflation problem 
(see \citealt{2010RPPh...73a6901B} 
for a review and the references therein). 

We performed a joint fit of the two G430L white light curves in
conjunction with the radial velocity data from
\cite{2009ApJ...693.1920H} 
and \cite{2011MNRAS.413.2500H}. 
We used the Markov Chain Monte Carlo (MCMC) suite {\tt EXOFAST} from
\cite{2013PASP..125...83E} 
to perform the joint fit.  In the joint fit, {\tt EXOFAST}
  utilises the \cite{2010A&ARv..18...67T} 
empirical polynomial
relations between the masses and radii of stars, and their surface
gravity, effective temperatures, and metallicities.  For the fit, we used
priors on the orbital period and central transit time from
section \ref{Sec:ephemeris} as well as priors on the stellar effective
temperature, gravity and metallically from \cite{2009ApJ...693.1920H}.
In addition, in order to incorporate the information contained within the G750L and WFC3
light curves, we also put priors on $a/R_{*}$ and
inclination based on the weighted-average transit fits to those datasets.
The results are given in Table \ref{Table:bestfit}.  In addition to
finding updated parameters, the {\tt EXOFAST} fits also serve to help verify
the Levenberg-Marquardt transit fits and errors, with all of the
fitted parameters and errors consistent at the 1-$\sigma$ level
between the two methods, when similar fits are performed, including
the values derived for the systematic error parameters described in
sections \ref{Sec:STISwhite} and \ref{Sec:wfc3white}.

The resulting planetary radius is in general somewhat larger than past
studies \citep{2011AJ....141..179C}, expected as the M-dwarf serves to dilute the transit depth.
With a joint fit using all of the available radial velocity data,
  our derived absolute radius is also nearly a factor of two more precise than recent ground-based
studies \citep{2013A&A...551A.108M}.  Consequently, our derived bulk
density of $\rho_P$=0.279$\pm$0.019 (cgs) for the planet is significantly lower than past studies (4.5-$\sigma$ compared to \citealt{2011AJ....141..179C}), further adding to the confirmation of an
anomalously large radius for this planet.

\begin{table} 
\caption{Best-fit joint parameters for WASP-12}
\label{Table:bestfit}
\begin{centering}
\renewcommand{\footnoterule}{}  
\begin{tabular}{lll}
\hline\hline  
Parameter & Description  & Value \\
\hline
\multicolumn{3}{c}{Stellar Parameters}\\
\hline
                           $M_{*}$\dotfill &Mass (\msun)\dotfill & $1.362_{-0.059}^{+0.060}$\\
                          $R_{*}$\dotfill &Radius (\rsun)\dotfill & $1.602_{-0.035}^{+0.037}$\\
                         $\rho_*$\dotfill &Density (cgs)\dotfill & $0.467\pm0.024$\\
             $\log(g_*)$$^\dagger$\dotfill &Surface gravity (cgs)\dotfill & $4.162\pm0.016$\\
               $\teff$$^\dagger$\dotfill &Effective temperature (K)\dotfill & $6400_{-180}^{+190}$\\
                             $\feh$$^\dagger$\dotfill &Metalicity\dotfill & $0.159\pm0.083$\\
\hline
\multicolumn{3}{c}{Planetary Parameters:}\\
\hline
                               $P$$^\dagger$\dotfill &Period (days)\dotfill & $1.09142166_{-0.00000031}^{+0.00000032}$\\
                        $a$\dotfill &Semi-major axis (AU)\dotfill & $0.02300_{-0.00034}^{+0.00033}$\\
                              $M_{P}$\dotfill &Mass (\mj)\dotfill & $1.417_{-0.048}^{+0.049}$\\
                            $R_{P}$\dotfill &Radius (\rj)\dotfill & $1.848_{-0.049}^{+0.052}$\\
                        $\rho_{P}$\dotfill &Density (cgs)\dotfill & $0.279\pm0.019$\\
                   $\log(g_{P})$\dotfill &Surface gravity\dotfill & $3.012_{-0.020}^{+0.019}$\\
            $T_{eq}$$^\ddagger$\dotfill &Equilibrium Temp. (K)\dotfill & $2578_{-73}^{+75}$\\
\hline
\multicolumn{3}{c}{RV Parameters:}\\
\hline
                    $K$\dotfill &RV semi-amplitude (m/s)\dotfill & $225.8_{-3.9}^{+3.8}$\\
                  $M_P\sin i$\dotfill &Minimum mass (\mj)\dotfill & $1.406_{-0.047}^{+0.048}$\\
                  $M_{P}/M_{*}$\dotfill &Mass ratio\dotfill & $0.000994\pm0.000022$\\
                $\gamma$\dotfill &Systemic velocity (m/s)\dotfill & $13.5\pm2.3$\\
\hline
\multicolumn{3}{c}{Transit Parameters:}\\
\hline
 $R_{P}/R_{*}$\dotfill &Radius of planet \dotfill & $0.11852_{-0.00080}^{+0.00081}$\\
      $a/R_{*}$$^\dagger$\dotfill &Semi-major axis\dotfill & $3.087_{-0.053}^{+0.052}$\\
                       $i$$^\dagger$\dotfill &Inclination (degrees)\dotfill & $82.72_{-0.72}^{+0.71}$\\
                            $b$\dotfill &Impact Parameter\dotfill & $0.391_{-0.033}^{+0.032}$\\
                          $\delta$\dotfill &Transit depth\dotfill & $0.01405\pm0.00019$\\
          $\tau$\dotfill &Ingress/egress dur. (days)\dotfill & $0.01534_{-0.00056}^{+0.00062}$\\
                  $T_{14}$\dotfill &Total duration (days)\dotfill & $0.1213_{-0.0012}^{+0.0013}$\\
         $u_1$\dotfill &linear limb-darkening coeff.\dotfill & $0.551_{-0.028}^{+0.029}$\\
      $u_2$\dotfill &quad. limb-darkening coeff.\dotfill & $0.220_{-0.021}^{+0.020}$\\
\hline
\end{tabular}
\end{centering}
\\
$^\dagger$parameter makes use of priors, see Sec. \ref{Sec:jointfit}
for details.\\
$^\ddagger$  $T_{eq}$ assumes zero albedo and a uniform planetary temperature
\end{table}

\begin{figure}
 {\centering
  \includegraphics[width=0.49\textwidth,angle=0]{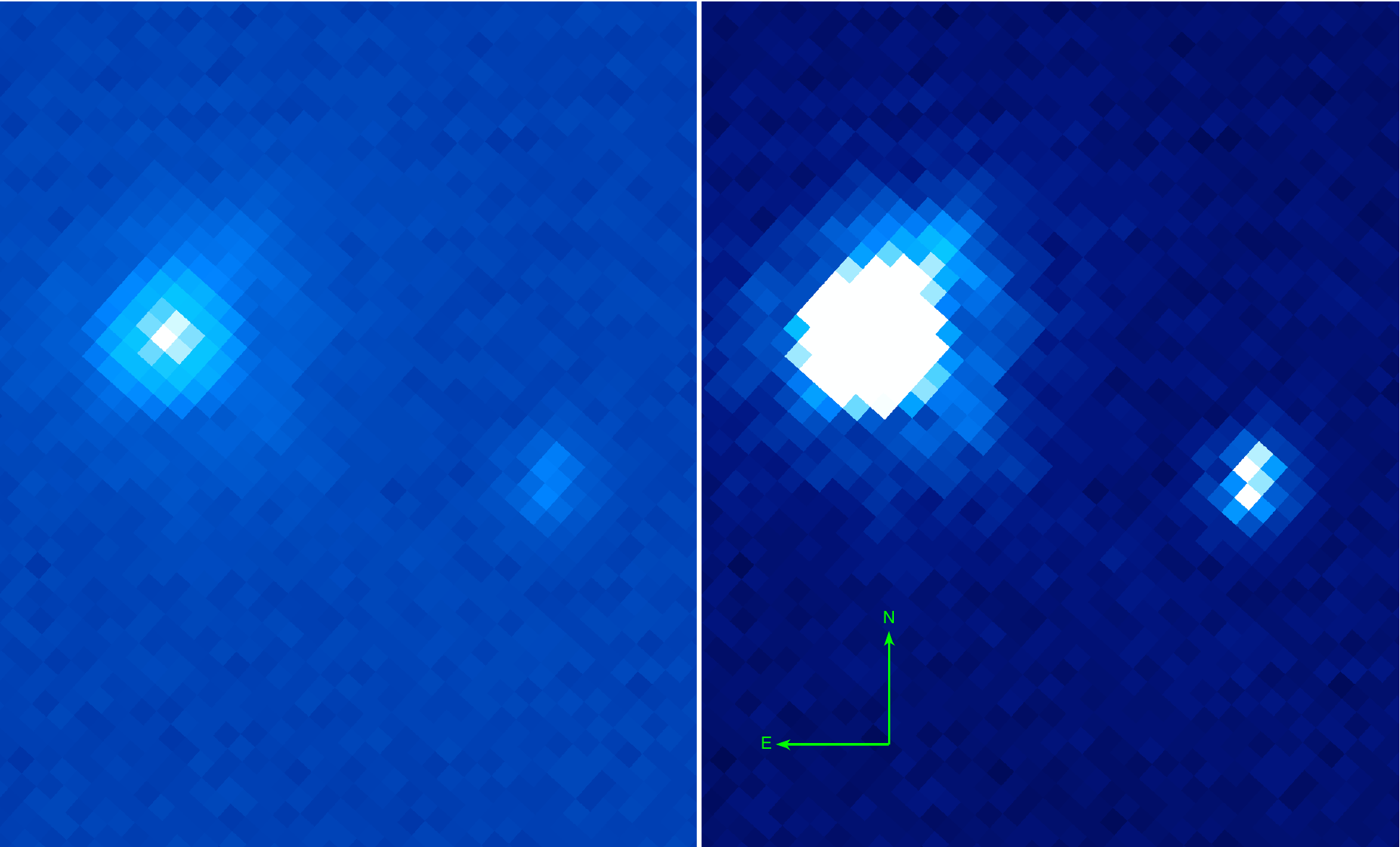}}
\caption[]{{\it HST} STIS optical acquisition image of WASP-12A and the two faint M-dwarf companions located 1.06$''$ to the south-west of the transit hosting star.  Shown is the image (left) on a logarithmic flux scale to illustrate the PSF of WASP-12A and (right) on a linear flux scale set to maximise the contrast of the north-south aligned M-dwarf pair.}
\label{Figure:acq}
\end{figure}

\subsection{M-dwarf stellar companions}
\label{Sec:Mdwarf}
The M-dwarf companion is well resolved in both the STIS acquisition
images (see Fig. \ref{Figure:acq}) and in the STIS spectra.  This
provides an opportunity to further characterise the properties of the
M-dwarf, and its impact on the WASP-12 system.
\cite{2012ApJ...760..140C} found that the M-dwarf was consistent with a
spectral type of M0V, with an effective temperature of  $T_{eff}$=3660~K using near-IR spectroscopy.  In addition, they found
that the system may
possibly be associated with WASP-12A if the M-dwarf was found to be
a close binary pair, but with only low resolution images at the time,
they could not firmly conclude its association.  However they did note
that the ground-based images had an extended PSF profile and that the two objects
had a common system radial velocity, pointing toward a binary M-dwarf pair.  

In order to align the {\it HST} PSF within the STIS longslit, an optical
broadband acquisition image was initially taken with the F28x50LP
filter, which has a wavelength range
between 5490 to 9990 \AA\ and a central wavelength of 7150 \AA.  With
{\it HST}'s high resolution, these images immediately reveal the M-dwarf to be a binary pair (see
Fig. \ref{Figure:acq}), confirming the suspicions of
\cite{2012ApJ...760..140C} and \cite{2013MNRAS.428..182B}.  
Further, the two components of the
M-dwarf binary are observed to have equal brightness, implying equal
mass, with a
flux contrast between the two estimated from the central pixel intensities to be 1.08$\pm$0.07 from the
acquisition image.  The 
  near unity flux contrast between the two stars also match the recent ground-based adaptive
  optics imaging results of \cite{2013arXiv1307.6857B}.  

\begin{figure}
 {\centering
  \includegraphics[width=0.49\textwidth,angle=0]{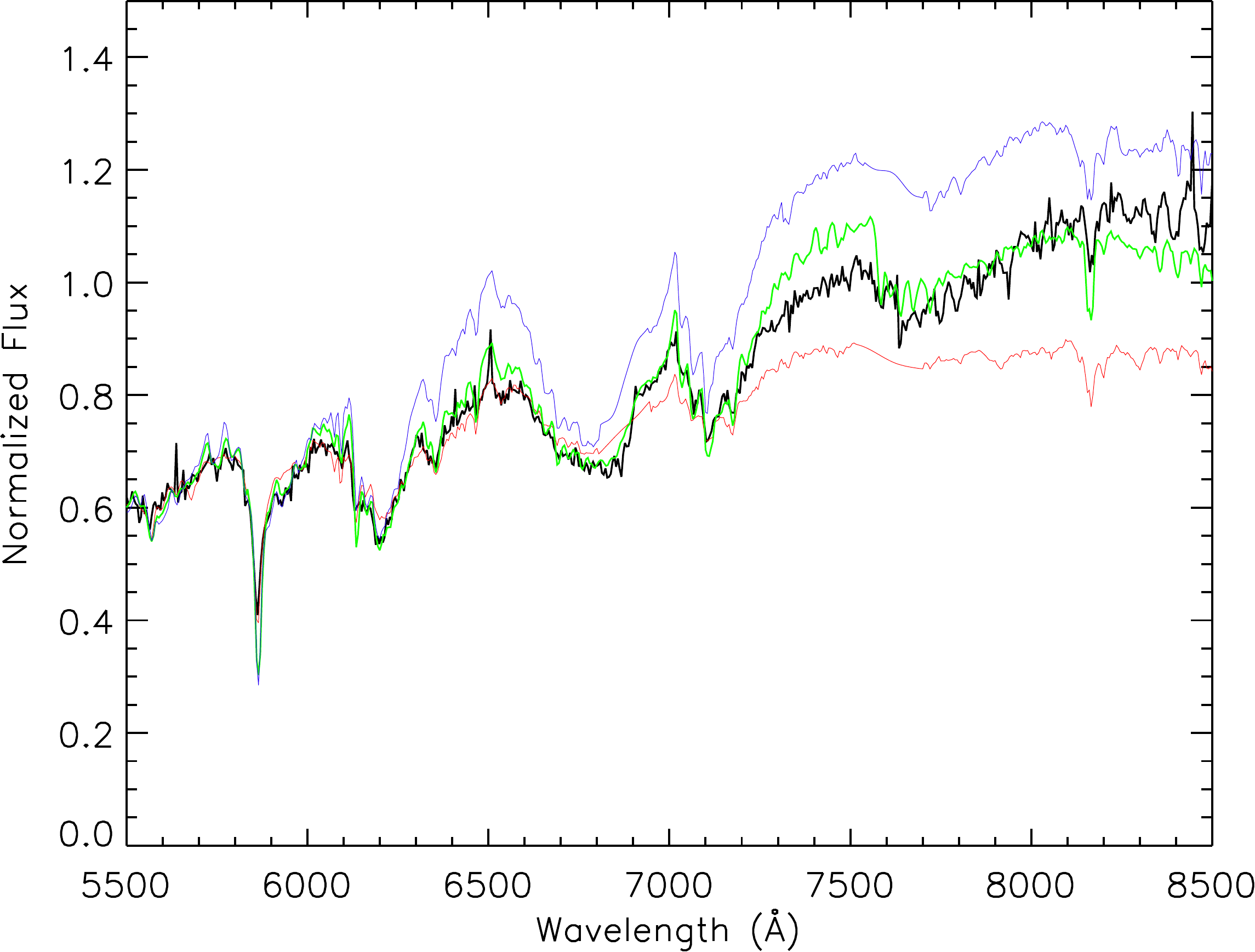}}
\caption[]{{\it HST} STIS G750L spectrum of the M-dwarf binary pair (black)
  compared to main sequence library stellar spectra taken from
  \cite{1998PASP..110..863P} 
 with $T_{eff}$ of 3800 (top blue) , 3680 (middle green) and 3550 (red
 bottom).  All spectra are normalised in flux at 5500~\AA.  The best
 matching 3680 K spectrum is consistent with the results of Crossfield
 et al. (2012), who determined a $T_{eff}$ of 3660$^{+85}_{-60}$ K
 corresponding to an M0V dwarf.  The STIS spectrum red-ward of about 7250
 \AA\ is affected by detector fringing, causing the observed deviations from the best-fitting template spectrum.}
\label{Figure:M0Vspec}
\end{figure}

Being spatially well resolved from WASP-12A in STIS spectral images, we also extracted the optical spectrum
of the M-dwarf companions (see Fig. \ref{Figure:M0Vspec}).  
The STIS G750L spectrum of the M-dwarf pair 
matches very well with the $T_{eff}$=3660~K measurement of \cite{2012ApJ...760..140C}, 
lending further evidence of the two stars possessing the same spectral
classification, with the spectral typing of the combined pair consistent
between the two studies from the optical to K-band.

Compared to WASP-12A, the flux contrast between the two M-dwarf stars and the
WASP-12A primary is 0.0207$\pm$0.0004 in the {\it HST} image, which is close to the $\Delta
i'$ measurement of \cite{2013MNRAS.428..182B}.  
Assuming an equal-mass M-dwarf binary, the distance modulus from \cite{2012ApJ...760..140C} 
is re-evaluated to be 7.85$\pm$0.2 mag for the M-dwarfs which compares
to 7.7$\pm$0.2 mag for WASP-12A, placing the three stars at the same distance
(within the errors), making a triple-star configuration highly likely.
At a distance of 427 pc \citep{2011AJ....141..179C}, 
the projected separation between WASP-12A and the M-dwarfs is 450 AU,
and the M-dwarf binary projected separation is about 40 AU.
A final proof of
association would be a common proper motion measurement between the
three stars (see \citealt{2013arXiv1307.6857B}).   

\subsection{Transmission spectrum fits}
\begin{figure*}
 {\centering
  \includegraphics[width=0.9\textwidth,angle=0]{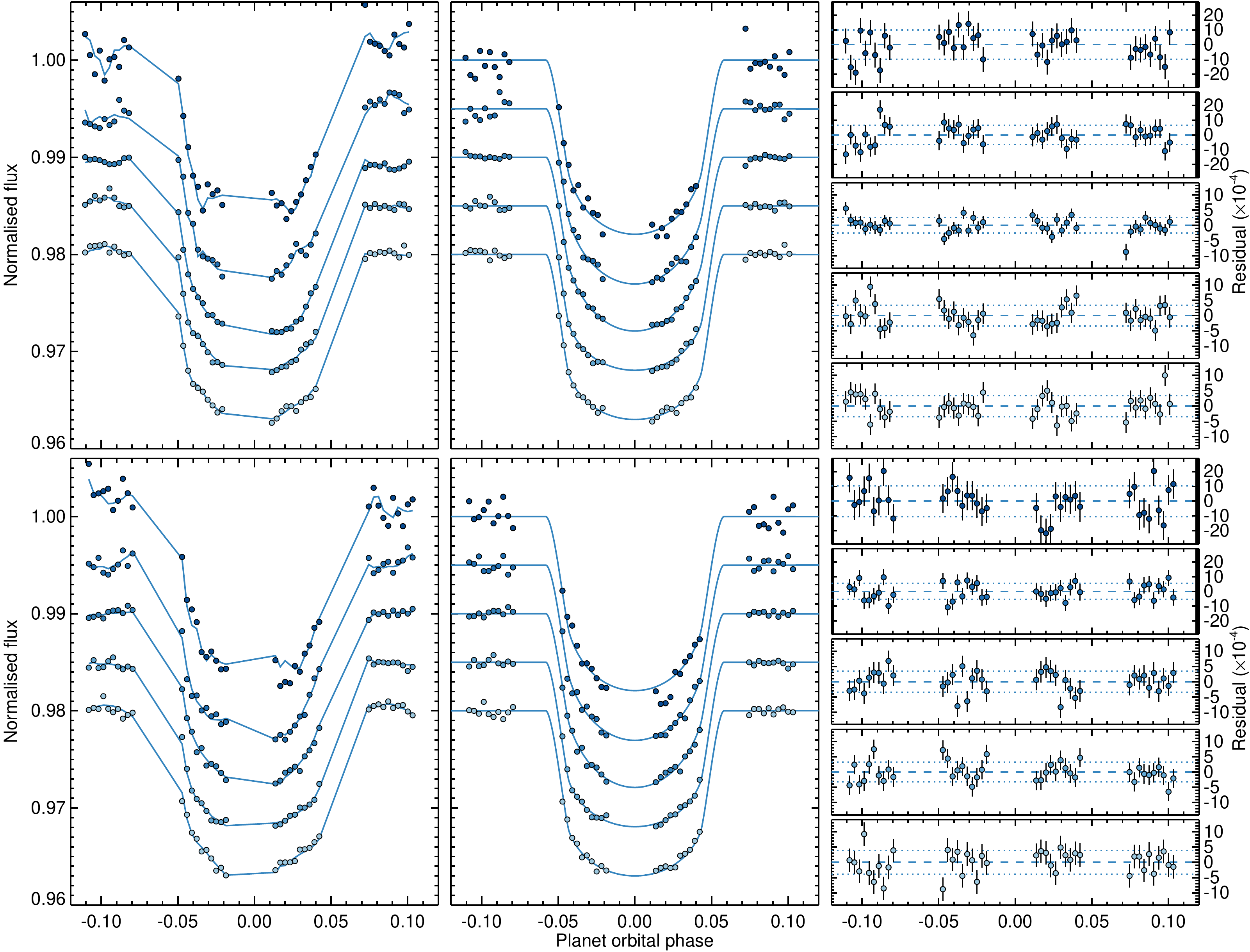}}
\caption[]{{\it HST} STIS G430L spectral bin transit light curves
  jointly fit for visit 1 (top) and visit 2 (bottom).  (Left) Light curves for the {\it HST} STIS G430L spectral
  bins with the common-mode trends removed, overplotted with the best fitting
  systematics model.  The points have been
  offset in relative flux, and systematics model plotted with connecting lines for clarity purposes.  The light curves are
  ordered in wavelength, with the shortest wavelength
  bin shown at the top and longest wavelength bin at the
  bottom. (Middle) Light curves fully corrected for systematic errors,
  with the best-fit joint transit model overplotted.  (Right) Residuals
  plotted with 1-$\sigma$ error bars along with the standard deviation
  (dotted lines).}
\label{Figure:specG430L}
\end{figure*}

\begin{figure*}
 {\centering
  \includegraphics[width=0.73\textwidth,angle=0]{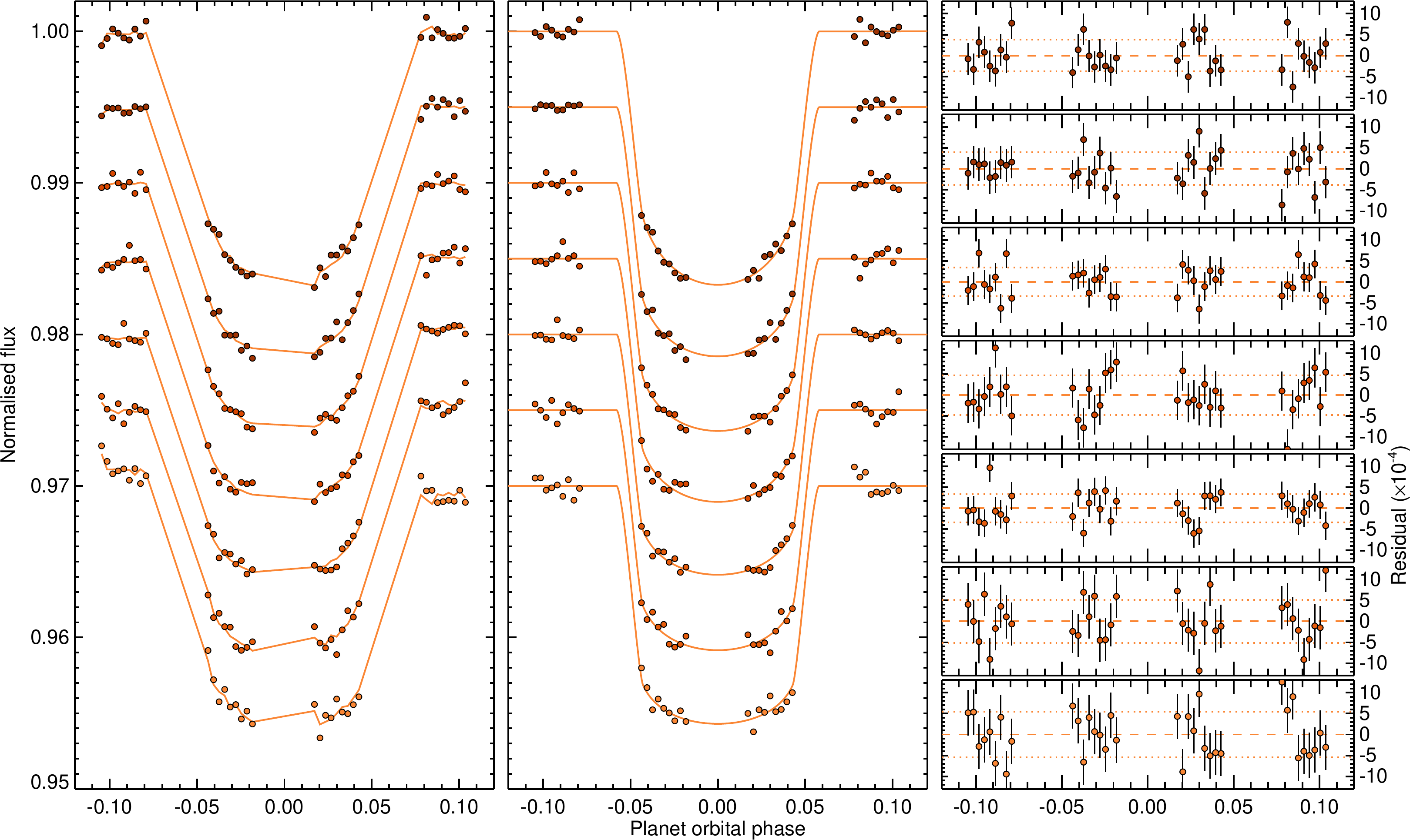}}
\caption[]{The same as Fig. 7, but for the G750L spectral bins.}
\label{Figure:specG750L}
\end{figure*}

\begin{figure*}
 {\centering
  \includegraphics[width=1\textwidth,angle=0]{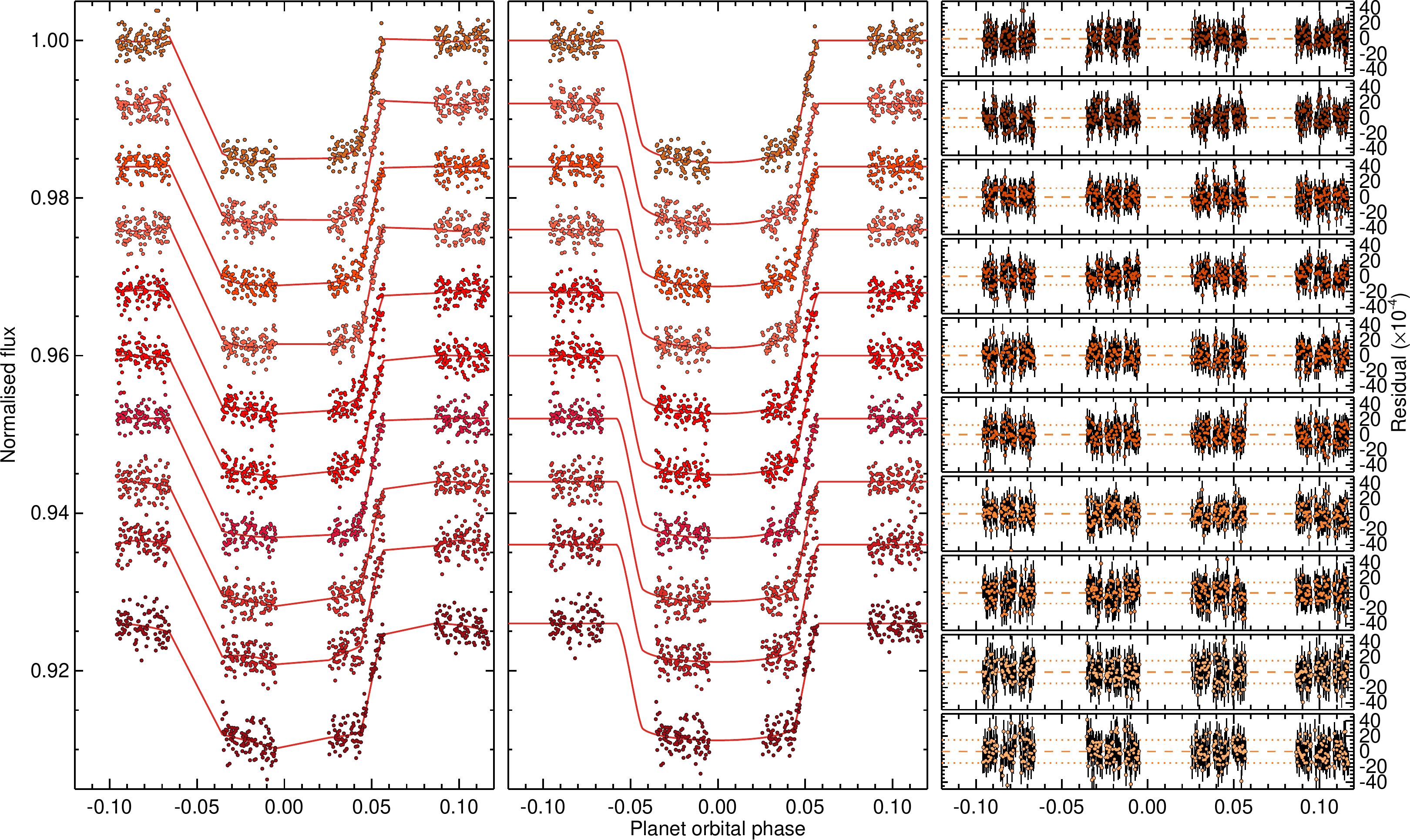}}
\caption[]{The same as Figs. \ref{Figure:specG430L} and \ref{Figure:specG750L}, but for the WFC3 G141
    spectral bins.}
\label{Figure:specWFC3}
\end{figure*}

We extracted various wavelength bins\footnote{available upon request} for the STIS G430L, G750L, and
WFC3 G141 spectra, to create a broadband transmission spectrum and
search for expected narrowband features (see
Figs. \ref{Figure:specG430L}, \ref{Figure:specG750L}, and  \ref{Figure:specWFC3}).  In these transit fits, we fixed the transit ephemeris to the results from
Sec. \ref{Sec:ephemeris} and fixed the inclination and stellar density
to their best-fitting values, allowing $R_{P}/R_{*}$ to be free as well as the
baseline flux and model parameters describing the instrument systematic trends.  
The WFC3 spectra were M-dwarf flux corrected in the same manner as the
white-light curve (see Sec. \ref{Sec:wfc3white}), with the uncertainty
from the correction well under each of our 1$\sigma$ errors for $R_{p}/R_{*}$. 

The limb-darkening parameters were 
fixed to the model values, with the four non-linear coefficients for each bin 
individually calculated from the stellar model, taking into account
the instrument response (see Table \ref{Table:LD}).
As a separate direct test of the limb-darkening model used, we
fit a light curve with a free limb
darkening coefficient in a spectral bin between 4000 to 4500 \AA\ using a
linear law, and compared the fit coefficient $u$ to the stellar model prediction.  At these wavelengths, the
predicted limb darkening is approximately linear (with a linear law deviating from ATLAS stellar model intensities by no more than 3.2\%, except at the very limb), making a
direct comparison of the stellar model and transit-fit limb-darkening coefficients
straightforward\footnote{Comparing multiple fit limb-darkening coefficients to the stellar models using higher
  order laws is difficult, as there are degeneracies between
  the fitted coefficients.
}.  The fitted linear coefficient is very close to the stellar model value,
with a simultaneous fit of the two G430L transits resulting in
$u=0.772\pm0.023$ which compares favourably to the ATLAS model
prediction of 0.774.  Further, we note that our transmission spectrum did not
  substantially change when fitting the light curves using a linear limb-darkening law with a
  freely fit coefficient, and the BIC favoured fixing the
  limb-darkening parameters to the model values.

We handled the modelling of systematic errors in the spectral bins by
two methods.  In the
first method, we allowed each light curve to be fit independently
with a parameterised model, as done in the white-light curve analysis.  In the second method, we removed the
common-mode trends from each spectral bin before fitting for the
residual trends, again with a parameterised model but fit with fewer parameters.
The common-mode trends were extracted from the white-light curves,
by removing the transits from the raw light curves using the parameters derived
from the best-fit models.  These common-mode trends were then removed from each
spectral bin, normalised to preserve the overall flux
levels.  In particular, removing the common trends reduces the amplitude of the
breathing systematics by a considerable margin, as these trends are
observed to be very similar wavelength-to-wavelength across the detector.

\begin{table} 
\caption{Non-linear limb darkening coefficients calculated for the
  G430L (top), G750L (middle) and WFC3 G141 (bottom).}
\label{Table:LD}
\begin{centering}
\renewcommand{\footnoterule}{}  
\begin{tabular}{ccccc}
\hline\hline  
 $\lambda$(\AA) & $c_1$ & $c_2$ & $c_3$ & $c_4$ \\
\hline  
  2900 - 3800     &0.2075 &  ~1.0926  & -0.4758 & ~0.0337  \\
  3800 - 4300     &0.2262  & ~0.8330  & -0.1838 & -0.0118  \\
  4300 - 4800     &0.5893  & -0.6147  & ~1.5498 & -0.6692  \\
  4800 - 5240     &0.3542  & ~0.7815  & -0.4770 & ~0.1080  \\
  5240 - 5700     &0.3816  & ~0.6909  & -0.3983 & ~0.0828  \\
\hline
  5300 -  5800    &0.3842  & 0.6811  & -0.3867 & ~0.0783  \\
  5800 -  6300    &0.4064  & 0.6433  & -0.4708 & ~0.1154  \\
  6300 -  6800    &0.4392  & 0.5138  & -0.3353 & ~0.0538  \\
  6800 -  7300    &0.4480  & 0.4646  & -0.2953 & ~0.0460  \\
  7300 -  7900    &0.4841  & 0.2891  & -0.1672 & ~0.0023  \\
  7900 -  8500    &0.4909  & 0.2526  & -0.1235 & -0.0147  \\
  ~8500 - 10300 &0.5009  & 0.1323  & -0.0197 & -0.0562  \\
\hline
11370 -  11890  & 0.4552  & 0.2941  & -0.3676 & ~0.1205  \\
11890 -  12410  & 0.4275  & 0.4063  & -0.4985 & ~0.1723  \\
12410 -  12930  & 0.4622  & 0.3591  & -0.5380 & ~0.2043  \\
12930 -  13450  & 0.4319  & 0.4980  & -0.7316 & ~0.2920  \\
13450 -  13970  & 0.4654  & 0.4057  & -0.6303 & ~0.2488  \\
13970 -  14490  & 0.5023  & 0.4198  & -0.7869 & ~0.3474  \\
14490 -  15010  & 0.5156  & 0.3447  & -0.6864 & ~0.3042  \\
15010 -  15530  & 0.6272  & 0.1071  & -0.5256 & ~0.2657  \\
15530 -  16050  & 0.6514  & 0.0618  & -0.5024 & ~0.2616  \\
16050 -  16570  & 0.6756  & 0.0228  & -0.4738 & ~0.2539  \\
\hline
\end{tabular}
\end{centering}
\end{table}

As found in \cite{Nikolov2013} for the HAT-P-1b STIS transit
data, we find that it makes little difference (i.e. deviations of $\sim$1$\sigma$ or less
on the fitted $R_{P}/R_{*}$ values) whether to include or not the position related
trends when fitting for the transit depths, as compared to
optimising the systematic trends for each bin individually based on
the BIC.  In the fully-parameterised method, we elected to fit the same model as used for the white-light curve
(4th order in {\it HST} phase and linear in X-position, Y-position, time, and
wavelength shift) as a some of the bins did prefer the model, and a uniform model for all wavelength
channels helps ensure that residual instrument trends are not differing
channel-to-channel, adversely effecting the transit depths.
With this method, we reach average precisions of 73\% the photon noise limit with the G430L, and 87\% for the
G750L, with no significant red-noise detected.

In the common-mode analysis, the spectral bins for the two G430L transits were each fit
with 3 fewer parameters compared to the fully-parameterised method,
as the 3rd and 4th order {\it HST} phase trends as well as the linear slope
did not justify being fit.  Similarly, for the G750L spectral bins,
the 4 orders of the {\it HST} phase
trend polynomial and wavelength shift trend did not need to be fit, reducing
the number of nuisance parameters for each light curve from 8 to 3.  
In addition, the common-mode analysis typically performed comparable or better than the
full-parameterised model, with the G430L attaining an average
precisions of 80\% the photon noise limit, and the G750L attaining
85\% the photon noise limit.  The biggest improvement is seen where the
grating efficiencies are low, as the spectral bins with lower counts are more
dominated by photon noise, making it harder for a parameterised model
to accurately fit for low-level systematic trends.

As done with STIS, we also fit the WFC3 spectral bins using a fully
parameterised model and a common-mode removal analysis.  As with the
STIS data, the common-mode trends derived from the WFC3 white light
curve dramatically reduced the apparent trends in the spectral bins.
 We report bin sizes which are comparable to the
STIS data, 560\AA\ wide, which is courser resolution than reported in
\cite{2013Icar..225..432S}  
as no obvious spectral features were apparent at
higher resolution.  

For the common-mode analysis, the number of
nuisance parameters was reduced from 5 to 2, with a quadratic function
of {\it HST} phase used to fit the residual trends.  With the fully-parameterised method, the precisions attained are close to
those of \cite{2013Icar..225..432S} 
who reported typical precisions near $\sim$86\% the photon noise limit.
The common-mode
analysis performed better, with an improvement of about 4\% in the residual scatter of the
binned light curves with the common-mode method.
For both reduction methods, when binning the data similarly as the two
aforementioned studies, we find nearly identical transmission spectral
results.  The one notable exception is the shortest wavelength channel of
Swain et al. (2012) which is more than 2$\sigma$ higher than our
results.  With the common-mode analysis, we also find that binning the
spectra to wider bandpasses does not significantly degrade the performance relative
to the photon limit, as found by \cite{2013Icar..225..432S} 
who adopted a linear analysis.

\begin{figure}
 {\centering
  \includegraphics[width=0.49\textwidth,angle=0]{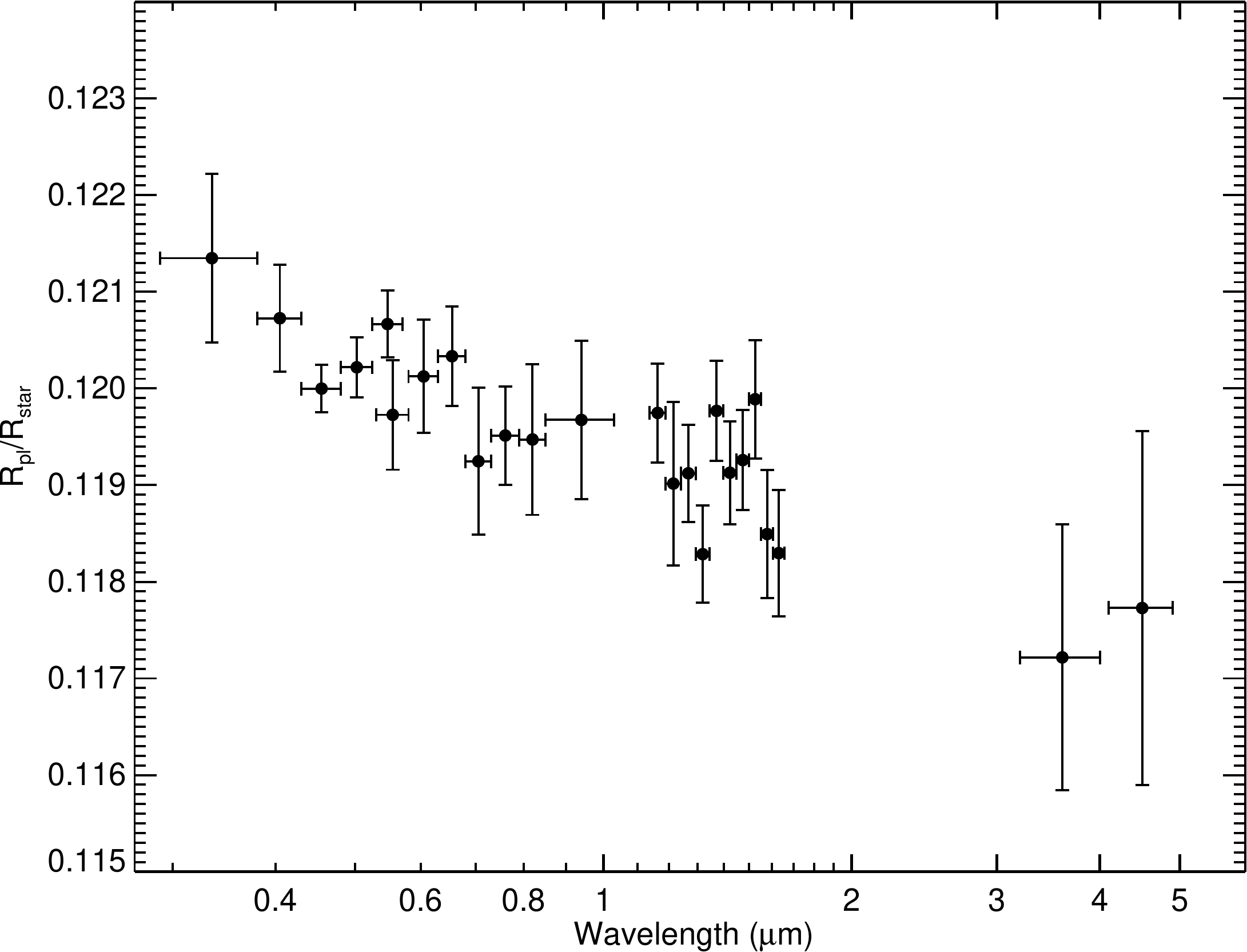}}
\caption[]{Combined {\it HST} STIS, {\it HST} WFC3 and {\it Spitzer} transmission
  spectrum of WASP-12b.}
\label{Figure:Broadbandspec}
\end{figure}

Given the overall better performance with significantly fewer nuisance parameters, we adopt
the common-mode analysis for further study, though note that all of the
subsequent results would be nearly identical if instead we adopted the
fully-parameterised analysis for both the STIS and
WFC3 spectra.  
The broadband spectral results are given in Table
\ref{Table:transmission} and shown in Fig. \ref{Figure:Broadbandspec},
where we have included the {\it Spitzer} transit photometry of
\cite{2012ApJ...747...82C} 
as re-evaluated with the M-dwarf dilution correction
by \cite{2012ApJ...760..140C}.  
We have chosen the non-prolate value for the 4.5 $\mu$m transit depth
measurement of \cite{2012ApJ...747...82C}, 
which is consistent with the WFC3 eclipse photometry of \cite{2013Icar..225..432S}. 

\begin{table} 
\caption{Broadband transmission spectral results for WASP-12b for the
  G430L (top), G750L (middle) and WFC3 G141 (bottom).}
\label{Table:transmission}
\begin{centering}
\renewcommand{\footnoterule}{}  
\begin{tabular}{cc}
\hline\hline  
 $\lambda$(\AA) & $R_{P}/R_{*}$  \\
\hline  
  2900 - 3800     & 0.12135  $\pm$   0.00087  \\
  3800 - 4300     & 0.12073  $\pm$   0.00056  \\
  4300 - 4800     & 0.12000  $\pm$   0.00025  \\
  4800 - 5240     & 0.12022  $\pm$   0.00031  \\
  5240 - 5700     & 0.12067  $\pm$   0.00035  \\
\hline
  5300 -  5800    & 0.11973 $\pm$    0.00057  \\
  5800 -  6300    & 0.12012  $\pm$   0.00059  \\
  6300 -  6800    & 0.12033  $\pm$   0.00051  \\
  6800 -  7300    & 0.11925  $\pm$   0.00076  \\
  7300 -  7900    & 0.11951   $\pm$  0.00051  \\
  7900 -  8500    & 0.11947  $\pm$   0.00078  \\
  ~8500 - 10300   & 0.11967  $\pm$   0.00082  \\
\hline
11370 -  11890  & 0.11975 $\pm$    0.00051  \\
11890 -  12410  & 0.11902  $\pm$   0.00085  \\
12410 -  12930  & 0.11912  $\pm$   0.00050  \\
12930 -  13450  & 0.11829  $\pm$   0.00050  \\
13450 -  13970  & 0.11977  $\pm$   0.00052  \\
13970 -  14490  & 0.11913  $\pm$   0.00053  \\
14490 -  15010  & 0.11926  $\pm$   0.00052  \\
15010 -  15530  & 0.11989  $\pm$   0.00061  \\
15530 -  16050  & 0.11849  $\pm$   0.00066  \\
16050 -  16570  & 0.11830  $\pm$   0.00065  \\
\hline 
36000 &                 0.11722  $\pm$   0.00138\\
45000 &                 0.11773  $\pm$   0.00183\\
\hline
\end{tabular}
\end{centering}

\end{table}


\section{DISCUSSION}

\subsection{Searching for narrowband spectral features}
In addition to deriving the broadband spectrum, we searched the STIS
data for narrowband absorption signatures, including the expected
species of Na, K, H$_{\alpha}$, and H$_{\beta}$, as the line cores of these
elements can have strong signatures.  In the context of the {\it HST} survey,
Na has been detected in Hat-P-1b by Nikolov et al. (2013),
who found the strongest signal in a 30\AA\ bandpass around the Na doublet.

We find no evidence for Na, either in wide or medium size bandpasses.
In a 30 \AA\ wide medium bandpass, we measure a planet-to-star radius difference of $\Delta
R_{p}/R_{*}=(-0.3\pm2.0)\times10^{-3}$ compared to a 600 \AA\ wide reference
region encompassing the Na doublet.  Theoretical models which are
assumed to be 
dominated by Na absorption in this region predict features of 4.0$\times10^{-3}$
$\Delta R_{p}/R_{*}$, so the data can rule out significant Na
absorption with about $95\%$ confidence.
Any Na feature present in WASP-12b must be confined to a narrow core,
similar to HD~189733b or XO-2b \citep{2012MNRAS.422.2477H, 2012MNRAS.426.1663S}. 

Similar to Na, we also find no evidence for H$_{\alpha}$, H$_{\beta}$,
nor K.  A 75 \AA\ wide bandpass
centred on the K doublet gives a radius difference of $\Delta
R_{p}/R_{*}=(-2.6\pm1.9)\times10^{-3}$.

\subsection{A thorough search for TiO}
As one of the most highly irradiated hot Jupiters discovered to date
with a zero-albedo equilibrium temperature of $T_{eq}$=2580 K, WASP-12b is a prime target to search for
signatures of TiO.  TiO has strong opacity throughout the optical,
with one of the strongest signatures being a `blue edge' at the short-wavelength end ($\sim$4300 \AA) of the molecule's optical bandheads, 
making the STIS G430L an ideal instrument to detect the species.  
Without the optical-blue wavelengths of STIS, ruling out
TiO in the atmosphere is not so straightforward.  For one, the
required optical-blue precisions are
difficult to obtain from the ground \citep{2013MNRAS.434..661C}.  
In addition, the overall higher planetary radius levels as seen in the optical-red
compared to the near-IR can be reasonably well matched by models which
include TiO, especially if the model abundances and temperature profiles
are fit \citep{2013arXiv1305.1670S}.  

\begin{figure}
 {\centering
  \includegraphics[width=0.49\textwidth,angle=0]{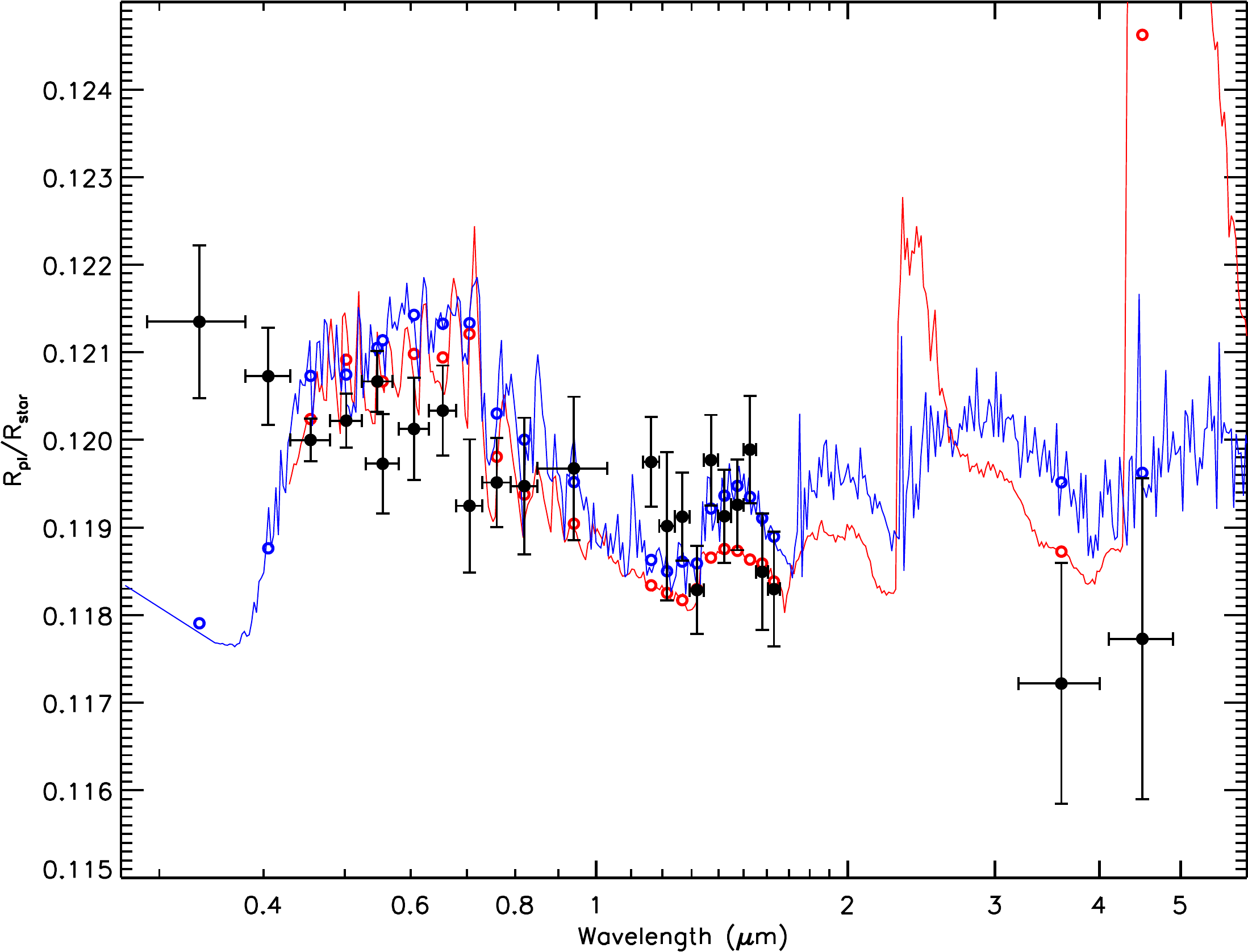}}
\caption[]{Broadband transmission spectra data of WASP-12b compared to
  atmospheric models containing TiO from (red) \cite{2010ApJ...719..341B}
and (blue) \cite{2010ApJ...709.1396F}.  
 The band-averaged model points are indicated with
  open circles.}
\label{Figure:TiO}
\end{figure}

Our broadband transmission spectrum shows no signs of TiO.  In
particular, while models expect the radius to drop by
2.4$\times10^{-3}$~$R_{p}/R_{*}$ blueward of 4300~\AA, our measured radii instead increase by an
average of (0.9$\pm0.5)\times10^{-3}$~$R_{p}/R_{*}$ compared to the 4300-4800 \AA\ bin,
representing a 6.6$\sigma$ confident non-detection of TiO at the `blue-edge'.
Radiative transfer models assuming equilibrium chemistry and solar
compositions have strong TiO features, and offer overall unsatisfactory fits to the 
broadband data, with a \cite{2010ApJ...709.1396F} 
model giving a $\chi^2$ of 44 for 23 DOF
(see Fig. \ref{Figure:TiO}).  

As our broadband spectrum is not optimally configured to detect
individual TiO bandheads, we employed custom tailored bins following
the TiO `comb' approach of 
\cite{2013arXiv1307.2083H} 
which sub-divides the region between 4840 and 7046 \AA\ into two optimal
bandpasses.  An `in' bandpass specifically covers wavelengths of 
8 strong TiO bandheads, while an `out' bandpass covers the wavelength regions
between the bandheads.  A positive TiO detection would show up as an
increased radius in the `in' bandpass compared to the `out', and a
positive [TiO]$_{\rm comb}$ value.  In this manner, we are able to use a very large
portion of the STIS data to search for specific TiO features,
providing good sensitivity.  This search also rules out TiO, with the
G430L and G750L giving an average [TiO]$_{\rm comb}$ radius difference of
$\Delta R_{p}/R_{*}=(-0.51\pm0.42)$$\times10^{-3}$ while WASP-12b models with TiO
predict a difference of +0.82$\times10^{-3}$~$R_{p}/R_{*}$, representing a 3.2$\sigma$ confident
non-detection of molecular bandheads.

As done in \cite{2013arXiv1307.2083H}, 
we also searched for the distinct `red edge' at 7396 \AA, where the broadband opacity
of TiO drops significantly toward longer wavelengths.  No significant difference in altitude was
found with a $\Delta R_{p}/R_{*}$ measured to be [TiO]$_{\rm red}=(-0.87\pm0.71)$$\times10^{-3}$  between an `in'
band from 6616-7396 \AA\ and an `out' band between 7396 and 8175
\AA, while models predict a difference of +1.4$\times10^{-3}$  $R_{p}/R_{*}$, a 3.2$\sigma$ confident
non-detection of the `red edge'.

With three independent tests to search for the molecule giving
sensitive null results, significant TiO absorption is decisively ruled out at high confidence as there are no spectral signs of the
characteristic `blue-edge', `red-edge', nor its strongest optical
bandheads.  Similar to Na or K, any TiO absorption features in the
  transmission spectrum of WASP-12b must be confined to narrowband
  signatures of the molecule's strongest optical transitions.

\subsection{A search for metal hydrides}
Metal hydrides, including TiH and CrH, have been suggested and explored as possible opacity
sources in WASP-12b when modelling the red-optical and near-IR
data \citep{2013Icar..225..432S, 
2013arXiv1305.1670S}, 
with TiH a potentially important Ti-bearing molecule in high C/O
  ratio atmospheres \citep{2012ApJ...758...36M, 2013arXiv1305.1670S}. 
A complete optical spectrum in this regard helps constrain the
presence of metal hydrides.  In particular, as TiH has its strongest bandhead near 5300~\AA\ 
\citep{2007ApJS..168..140S}, 
the G430L is very sensitive to its presence.  

We find no evidence for
TiH, with narrower bandpasses around 5300~\AA\ all consistent
with the broadband level near 0.1207 $R_{p}/R_{*}$.  The lack of the strong
5300~\AA\ TiH bandhead constrains the presence other longer-wavelength 
bandhead features of TiH to be at or below the observed optical and
near-IR radius.  Thus, we conclude TiH can not be a major
broadband opacity source for the transmission spectra.  The same is
true for MgH, which also has its strongest peak in the optical-blue \citep{2007ApJS..168..140S}, which we also do not observe.

Significant opacity by CrH can also be excluded, as abundances needed
to match the WFC3 near-IR $R_{p}/R_{*}$ levels would produce strong features 
between 8000 and 10,000~\AA\ which would rise to $\sim$0.122 $R_{p}/R_{*}$, where the opacity of CrH is strongest
\citep{2007ApJS..168..140S}, though no such features are apparent.

In addition to these molecule-specific searches, we also explored a
range of different atmospheric models with varying opacity (0.1 to 10
solar) as calculated in \cite{2010ApJ...719..341B}.  
These models contained significant metal hydride features (including
MgH, FeH, and CaH), though
none were satisfactory fits to the broadband data (see Table \ref{Table:fits}).

\subsection{Limits on molecular absorption}
Despite different reduction techniques and instrument systematic
models, the shape of our near-IR transmission spectra agrees very well with other
studies \citep{2013Icar..225..432S, 
2013arXiv1305.1670S}.  
The data is of sufficient quality to have detected H$_{2}$O, if it
were present at the levels some models predict, with the WFC3 atmospheric H$_{2}$O detections
in WASP-19b and HAT-P-1b a direct example \citep{2013arXiv1307.2083H, 2013arXiv1308.2106W}. 
In agreement with the findings of \cite{2013Icar..225..432S}, 
we find no strong evidence for H$_{2}$O absorption, though there is a
small `bump' at 1.4 $\mu$m which is compatible with a small amplitude H$_{2}$O
feature.  A straight-line (with 2 free parameters fit in
$R _{p} (\lambda)/R_{*}$ vs. ln$\lambda$) remains a good fit to the WFC3 data ($\chi^2=9$ for 8 DOF).  

A small amplitude H$_2$O feature would imply that either the
scale height is smaller than expected (requiring lower temperatures),
the H$_2$O abundance is low, or that the feature is being covered by other absorbers (including hazes or
clouds, also see section \ref{section:partlycloudy}) as has been
suggested for HD209458b \citep{2013ApJ...774...95D}. 
We find that the WASP-12b WFC3 transmission spectrum is compatible with H$_{2}$O-only
absorption if the temperature is below 1600 K at 95\% confidence.
This temperature presents a problem when interpreting the overall broadband spectrum, as 
smaller scale heights from lower temperatures would lead to a flatter
overall spectrum, yet significant broadband radii differences are observed.
In addition, as pointed out by \cite{2013arXiv1305.1670S},  
the {\it Spitzer} data is compatible with smaller radii,
even though models with significant H$_{2}$O (and CO) absorption predict
larger radii at those longer wavelengths.

Potentially, HCN could be an important molecule in the
atmosphere, especially if WASP-12b were to have a high C/O ratio
\citep{2013ApJ...763...25M, 2013arXiv1307.2565B}.
However, the cross section of HCN is considerably weaker in the
optical and near-IR compared to {\it Spitzer} wavelengths (e.g. see \citealt{2011ApJ...727...65S}), which places  
constraints in our broadband spectrum.
As found by \cite{2013arXiv1305.1670S}, the near-IR and {\it Spitzer}
$R_{p}/R_{*}$ radius levels long-wards of 1.55$\mu$m can reasonably
reproduced assuming HCN dominates that region, but with a sharp drop in the
opacity short-ward of 1.55$\mu$m, our optical data and the
near-IR WFC3 up to that wavelength would need an additional opacity source to explain the
larger radius levels.  While \cite{2013arXiv1305.1670S} attributes the
higher optical levels in part to TiO or TiH, both molecules are
inconsistent with our blue-optical data.  With current
data, a re-analysis of the secondary eclipse data,
taking into account the M0V dilution effects, would likely place stronger
overall constraints on the presence of HCN than our transmission spectra.

\begin{figure}
 {\centering
  \includegraphics[width=0.49\textwidth,angle=0]{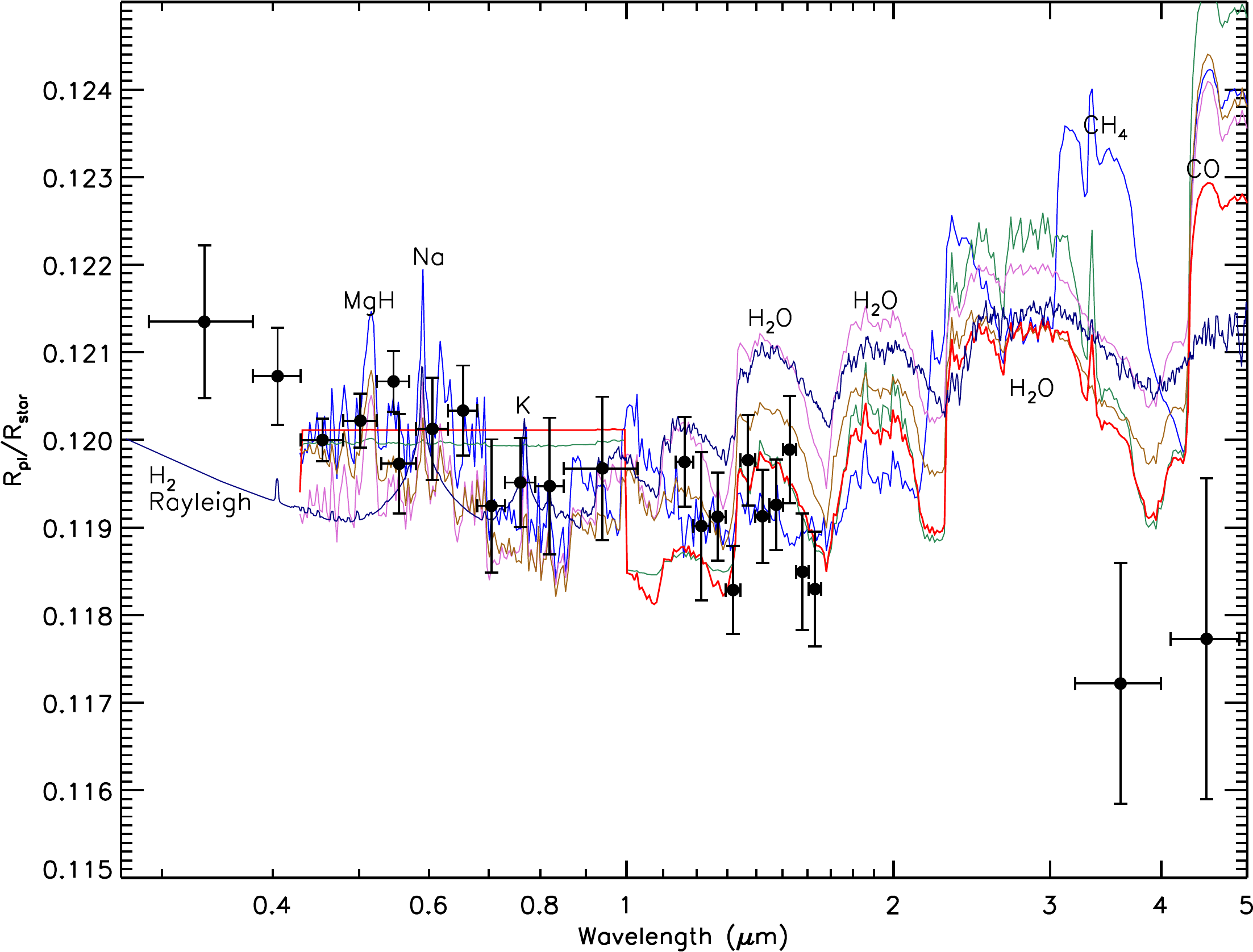}}
\caption[]{Plotted is the broadband transmission spectral data
  compared to 6 different clear atmosphere models (which lack TiO)
  listed in Table \ref{Table:fits}, including Burrows-ExtraAbsorber\_10$\times$solar
  (red), Burrows-MetalHydrides\_0.01$\times$H2O (light blue),
  Burrows-ExtraAbsorber\_10$\times$CO (green), Burrows-Isothermal3000\_0.1$\times$solar (brown),
  Burrows-Isothermal2500  (orchid), and Fortney-Isothermal2250\_noTiO (dark blue).
  All of these models have a particularly hard time simultaneously
  fitting for the near-IR WFC3 and {\it Spitzer}
  data.}
\label{Figure:Molecule}
\end{figure}

We compared our broadband spectrum to a variety of different clear atmospheric
models (dominated by gaseous molecular and atomic features) from the
1D modelling suites of \cite{2010ApJ...719..341B, 2012ApJ...756..176H} and \cite{2008ApJ...678.1419F, 2010ApJ...709.1396F}.
The models include T-P profiles assumed to be isothermal, as well as
day-side and planet-wide average profiles in radiative hydrostatic
equilibrium.  Models were also computed with a variety of different
metalicities, C/O ratios, and heat redistribution efficiency parameters as well as extra optical absorbers used by
\cite{2007ApJ...668L.171B} 
to produce stratospheres and interpret {\it Spitzer} secondary eclipse
data.  Models were also run with and without including gaseous TiO and
with/without including metal hydrides.  As shown in Table
\ref{Table:fits}, none of the models are particularly good fits to the
transmission spectrum data (see Fig. \ref{Figure:Molecule}).


\subsection{Interpreting the broadband transmission spectrum with aerosols}
While most of the expected molecular and atomic features can be
confidently ruled out (e.g. TiO, Na, \& K) as major contributors to our broadband
  transmission spectrum, the overall slope of the 
spectrum shows a significant broadband atmospheric feature, with an overall characteristic slope which spans
0.0031$R_p/R_{*}$ from the optical to the infrared (see
Fig. \ref{Figure:Broadbandspec}).  The atmospheric pressure scale
height $H$ is given by $H=\frac{kT}{\mu g}$, where $k$ is Boltzmann's
constant, $T$ is the local gas temperature, $\mu$ is the mean molecular
weight of the atmosphere, and $g$ is the surface gravity.
For WASP-12b, the scale height is estimated to be 700 km at 2100 K, implying that our broadband 
transmission spectrum spans $\sim$5$H$ in altitude (3500 km).

Following \cite{2008A&A...481L..83L}, we
fit the broadband transmission spectrum assuming an atmospheric opacity source(s)
with an effective extinction (scattering+absorption) cross section which follows a power law
of index $\alpha$, i.e. $\sigma=\sigma(\lambda/\lambda_0)^{\alpha}$.
With this assumption, the slope of the transmission spectrum is then proportional to the
product $\alpha T$ given by,
\begin{equation}  
  \alpha T=\frac{\mu g}{k} \frac{{\rm d}(R_{p}/R_{*})}{{\rm d ln}\lambda}. 
\end{equation}
With these assumptions, we find a good fit to the 24 data points fitting for
$\alpha T$ and the baseline radius,  ($\chi^2$ of 16.8 for 22 DOF, $N_{free}=2$)
resulting in $\alpha T= -3528\pm$660 K.

Assuming a temperature mid-way between the day-side and night-side
temperatures of 3000 K and 1100 K respectively\footnote{The day and night temperatures, $T_{day}$ and $T_{night}$, are calculated using the bond albedo
and recirculation efficiency from Cowan et al. (2012), the definitions from Cowan et al. (2011), and the parameters in Table \ref{Table:bestfit}.} and allowing a temperature range
which encompasses a $T_{eq}$ of 2578 K, gives a temperature in the range of
2100$\pm$500 K.  With these adopted temperatures,
the slope of the transmission spectrum would imply an effective
extinction cross section of
$\sigma=\sigma(\lambda/\lambda_0)^{-1.7\pm0.5}$.

\subsubsection{Rayleigh scattering}
\label{section:Rayleigh}
Assuming that the atmosphere can be described by Rayleigh scattering
($\alpha=-4$), as would be the case in a pure H$_2$ atmosphere or one
for an atmosphere dominated by a highly scattering aerosol 
as observed in HD~189733b 
\citep{2008MNRAS.385..109P, 2008A&A...481L..83L, 2011MNRAS.416.1443S}, 
the slope would imply temperatures of 882$\pm$165 K.  
In the H$_2$ case, the fit gives a reference pressure of
$550^{+470}_{-170}$ millibar at $R_{p}/R_{*}$ = 0.1195.
This reference pressure is an upper limit representing a
clear atmosphere, with any other assumed
opacity source for the broadband spectra necessarily residing at lower pressures.
With the Rayleigh scattering assumption, the low fitted
temperatures serve to reduce the scale height, and decrease the
modelled transmission spectral slope to match the data.
While this temperature is consistent with the night-side brightness
temperature found by \cite{2012ApJ...747...82C}, 
it is perhaps too low to be representative of the limb, which likely
has intermediate temperatures between those of the night and
day-side.  Such moderate limb temperatures are supported by 3D 
models, with \cite{2010ApJ...709.1396F} 
finding that planet-wide average $T-P$ profiles are a good approximation
to the true limb profiles output from 3D simulations.

\subsubsection{Mie scattering dust}

While Rayleigh scattering can not be excluded, especially if the
high-altitude temperatures at the exoplanetary limb are much colder
than anticipated, good fits to the data are also possible assuming
significant opacity from aerosols. 
In particular, materials with significant absorptive
properties (i.e. a non-negligible complex component to the index of
refraction) can fit the
observed slope assuming Mie theory.  In general, such aerosols can be
produced from condensate dust species, or alternatively from photochemistry.
For the hot day-side temperatures measured for WASP-12b, 
high temperature dust condensates such as corundum (Al$_2$O$_3$), iron oxide (Fe$_2$O$_3$), or perovskite (CaTiO$_3$) are candidate materials. 
\begin{figure*}
 {\centering
  \includegraphics[width=0.9\textwidth,angle=0]{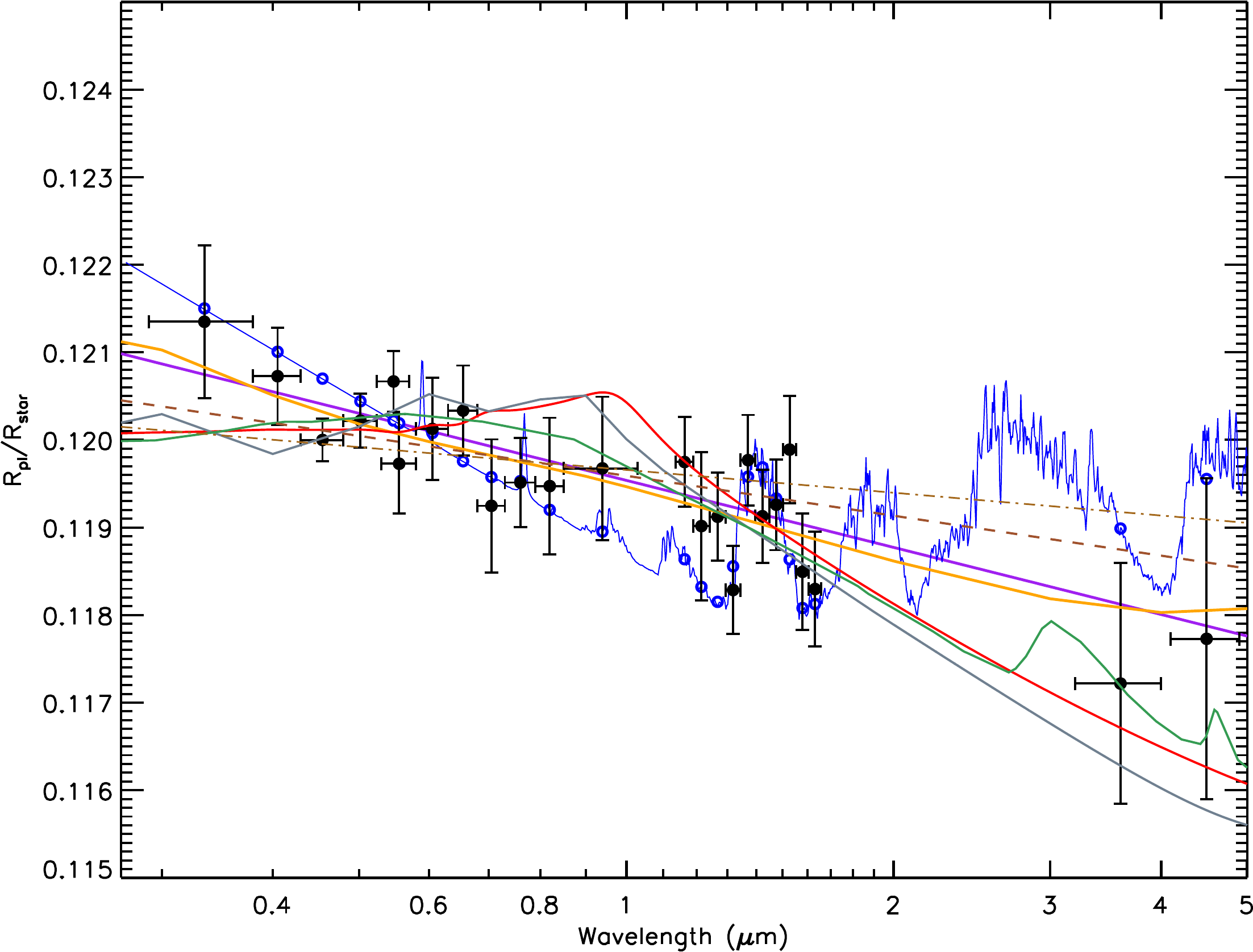}}
\caption[]{Plotted is the broadband transmission spectral data
  compared to 6 different aerosol models listed in Table 4 including:
  Rayleigh scattering (purple), Mie scattering Al$_2$O$_3$ (orange), Settling
  dust $\beta$$=-3$ (brown dashed),  Mie scattering Fe$_2$O$_3$ (red),
  Mie Scattering CaTiO$_3$ (gray),
  Settling Dust $\beta$$=-0$ (light brown dot-dashed), a partly cloudy
  model with enhanced Rayleigh scattering Fortney-noTiO\_EnhancedRayleigh
  (blue), and Titan tholin (green).}
\label{Figure:Dust}
\end{figure*}

We tested various atmospheric models assuming aerosol dust was the dominant
spectral feature at all observed wavelengths.  
We compiled the optical properties
of Al$_2$O$_3$ \citep{1995Icar..114..203K},
Fe$_2$O$_3$
(Triaud, A.\footnote{http://www.astro.uni-jena.de/Laboratory/OCDB/oxsul.html}),
and CaTiO$_3$ (J. W. Ferguson, private communication).
Using Mie theory, we calculated the
extinction cross sections as a function of wavelength for a large range of
particle sizes, $a$, then used the analytical formula of
\cite{2008A&A...481L..83L} to calculate the wavelength-dependant
transmission spectrum.  As the cross section is dominated by the
largest grain size in a distribution with $\sigma \propto a^6$
\citep{2008A&A...481L..83L}, 
the transmission spectrum can be reasonably fit to first order simply assuming a single grain size.  

In fitting Mie scattering dust models against the transmission spectrum, we found a degeneracy
between the grain size and atmospheric temperatures, which is expected
following equation 2,
as $a$ is related to $\alpha$ through the cross-section, and both
$\alpha$ and $T$ have an influence on the resulting slope.  In
particular for models assuming Fe$_2$O$_3$, 
when $a$, $T$, and the baseline altitude were all set to freely fit
the data, the resulting fits favoured sub-micron particle sizes with
low temperatures near
1000~K required, resulting in fits quite similar to the Rayleigh
solution.  
In addition, very small MgSiO$_3$ and CaTiO$_3$ particles have scattering properties within
the Rayleigh regime, and thus also give good fits near 1000~K equivalent to those given in section \ref{section:Rayleigh}.
A similar fit for Al$_2$O$_3$ ($\chi^2$ of 16.7 for 21 DOF) resulted in somewhat warmer temperatures
of 1400$\pm$400 K.

To explore the validity of high temperature condenses at the expected
temperature regime, we also fit Mie scattering models fixing the
$T$ to a value of 2100 K and fitting for
$a$ and the baseline radius.  2100 K
temperatures are not only plausible for the limb of WASP-12b, but are also
near the condensation temperatures for the dust species in question.
Fe$_2$O$_3$ 
results in a somewhat less satisfactory fit ($\chi^2$ of 24.8 
respectively for 22 DOF) as it contains optical features which are not
apparent in our data which become pronounced at warmer temperatures
(see Fig. \ref{Figure:Dust}).
As CaTiO$_3$ is highly scattering throughout the optical, it requires
  particle sizes near 0.18$\mu$m 
  ($\chi^2$ of 24.0
respectively for 22 DOF),  which produces a fit similar in quality as
Fe$_2$O$_3$.  
For the dust species tested, Al$_2$O$_3$ was the best fit against the transmission spectrum
($\chi^2$ of 16.8 for 22 DOF).  As Al$_2$O$_3$ does not have strong
scattering or absorption features over our wavelength range, with the
$a$-$T$ degeneracy, for sizes below $\sim$0.1$\mu$m equally good fits can be found at low, moderate, and high temperatures
(e.g. 1000, 2100, 3000 K).  
For Al$_2$O$_3$, grain abundances and corresponding atmospheric
pressures for the transmission spectrum are estimated using solar abundances
and assuming the condensate is limited by the available number of aluminium atoms.
We find corresponding pressures of 0.02 and 0.5 millibar at $R_{p}/R_{*}$ =
0.1195 if the particle sizes are 0.3 and 0.03 $\mu$m respectively.   
For all assumed temperatures greater than
1450 K, Al$_2$O$_3$ remains our best fitting model.

\subsubsection{Settling dust}

\cite{2013MNRAS.432.2917P} 
explored the case where the atmospheric opacity is dominated by Rayleigh scattering
from grains of condensates, with the density and
maximum size of the grains varying with pressure (“settling grains”).
In this case, the slope of the transmission spectrum, in units of
altitude $z$
per ln$\lambda$, was found to be $-\frac{4}{7}H$ in the case of a flat distribution
of grain sizes ($\beta = 0$), and $-1H$ in the case of an equal partition
of mass across all grain sizes ($\beta = -3$), where $\beta$ is a
power law index relating the abundance at height $z$ in the atmosphere
to the particle size.
With a scale height assuming a 2100 K temperature, the $\beta = 0$
case can be excluded from the data as it results in a shallower slope
than observed  ($\chi^2$ of 29.3 for 22 DOF), though the $\beta = -3$
case results in a better fit ($\chi^2$ of 21.5 for 22 DOF).
Allowing the temperature to freely fit can also reproduce a
fit equally well as Rayleigh scattering (as it has the effect of increasing the slope), but requires temperatures 
of 3528 K in even the $\beta = -3$ case.

\subsubsection{Partly cloudy}
\label{section:partlycloudy}

Finally, we explored fitting the data to a suite of
\cite{2010ApJ...709.1396F} models which had either an artificially added Rayleigh
scattering component simulating a scattering haze, or a `cloud deck' simulated by a grey flat line
at different altitudes.  These models serve to test the hypotheses
that clouds or hazes help cover some, but not all, of the atomic and molecular
atmospheric features expected.  A 2100 K model lacking TiO with a
Rayleigh scattering component with a cross section 100$\times$ that of H$_2$, resulted in a fit with $\chi^2$ of 31.6 for 22 DOF.

\subsubsection{Tholin haze}
\label{section:tholins}

We tested Mie scattering models assuming that a hydrocarbon haze was the dominant
spectral feature, an important feature in solar system planets and a
possibility for giant exoplanet atmospheres
\citep{2003ApJ...588.1121S}. 
Tholins have a characteristic red colour, as they preferentially
absorb at shorter visible wavelengths.
We tested the optical properties of a Titan tholin haze from
\cite{1984Icar...60..127K} 
measured in a N$_2$ atmosphere, as well as a H$_2$-atmosphere tholin
haze from \cite{1987JGR....9215067K}, 
whose optical properties are only available between 0.4 to 1 $\mu$m and has weaker overall absorptivity.
Both tholins absorb strongly in the blue, with similar Mie scattering extinction
cross sections which are close to
$\sigma=\sigma(\lambda/\lambda_0)^{-4}$ between 0.3 and 2~$\mu$m
for particles $<0.1\mu$m (also see \citealt{2012ApJ...756..176H}).
Similar to the Fe$_2$O$_3$ 
fit, the Titan and H$_2$ tholin scattering fits with $a$, $T$, and the baseline radius free
were similar to the Rayleigh solution, as the lower $\sim$1000 K
temperatures served to mute the tholin's strong blue absorption through a decreased scale height.  Fixing the temperature to
2100 K, the tholins provided satisfactory fits the transmission spectral data (see Table \ref{Table:fits}) with particle sizes
near 0.25$\mu$m, which is large enough such that cross section becomes
fairly flat over most of the optical and only decreases redward of
1$\mu$m (see green model in Fig. \ref{Figure:Dust}).  We note that Jupiter has similarly sized aerosols in its
stratosphere at low latitudes (\citealt{Zhang2013159}).  At 2100 K, particle sizes smaller than 0.19$\mu$m are ruled
out in WASP-12b, as the predicted slope becomes much too steep to fit the data.

\subsection{The case for aerosols in WASP-12b's atmosphere}

\begin{table*} 
\caption{Model selection fit statistics against the $N=24$ broadband
  transmission spectrum datapoints. Atmospheric models including aerosols
  are listed at the top, while clear atmospheric models (dominated by
  gaseous molecular and atomic absorption features) are listed at the
  bottom. The $\Delta$BIC and relative probability for all models is calculated with respect to the
  Rayleigh scattering model.  The Burrows et al. (2010) models
are compared against the 22 longest wavelength data points, while the
Fortney et al. (2010) models and are compared against all 24 points.
 }
\label{Table:fits}
\begin{centering}
\renewcommand{\footnoterule}{}  
\begin{tabular}{lccccc}
\hline\hline  
Models (with aerosols)                     &$\chi^2$& $N$, DOF  & BIC & D=$\Delta$BIC/2 & Prob=exp(D)\\
\hline  
Rayleigh scattering                   & 16.8      & 24, 22     & 23.1 & 0        &  1\\
Mie scattering Al$_2$O$_3$      & 16.8      & 24, 22     & 23.1  & 0 & 1  \\
Titan tholin                               & 20        & 24, 22      & 26.4& -1.7 & 0.19 \\   
Settling Dust ($\beta=-3)$       & 21.5      & 24, 22     & 27.9 & -2.4   &  0.09 \\
Mie scattering CaTiO$_3$          & 24.0      & 24, 22     & 30.3  & -3.6  & 0.03  \\
Mie scattering Fe$_2$O$_3$       & 24.8      & 24, 22     & 31.2  & -4.0  & 0.02  \\
Settling Dust ($\beta=0)$         & 29.3      & 24, 22     & 35.7  &
-6.3      &  0.002 \\

Fortney-noTiO\_EnhancedRayleigh    & 31.6      & 24, 22    & 40 & -7.4   &  0.0006 \\
Fortney-noTiO\_CloudDeck  & 47.4      & 24, 22    & 54   & -15    & 10$^{-7}$  \\
\hline
Models (clear atmospheres)        &$\chi^2$& $N$, DOF  & BIC & D=$\Delta$BIC/2 & Prob=exp(D)\\
\hline
Burrows-ExtraAbsorber\_10$\times$solar         &28.5       & 22, 20    &  34.7   &-6.3    & 0.0018\\
Burrows-MetalHydrides\_0.01$\times$H2O  &39     &22, 21      & 45    &-12     &10$^{-5}$\\
Burrows-ExtraAbsorber\_10$\times$CO &35        &22, 19     & 44    & -11    &10$^{-5}$\\
Fortney-Isothermal2250\_withTiO          & 44        & 24, 23    & 48   & -12 & 10$^{-5}$  \\
Burrows-Isothermal3000\_0.1$\times$solar           &54       &22, 20      & 60     & -19  & 10$^{-8}$ \\
Burrows-withTiO                      &58       &22, 21      & 61  &-19      &10$^{-9}$\\
Burrows-Isothermal2500                                  &115     &22, 21      &   118 &-48   & 10$^{-21}$ \\
Fortney-Isothermal2250\_noTiO                   & 134      & 24, 23     &137     & -57  &10$^{-25}$  \\

\hline
\hline
\end{tabular}
\end{centering}

\end{table*}

\begin{figure}
 {\centering
  \includegraphics[width=0.49\textwidth,angle=0]{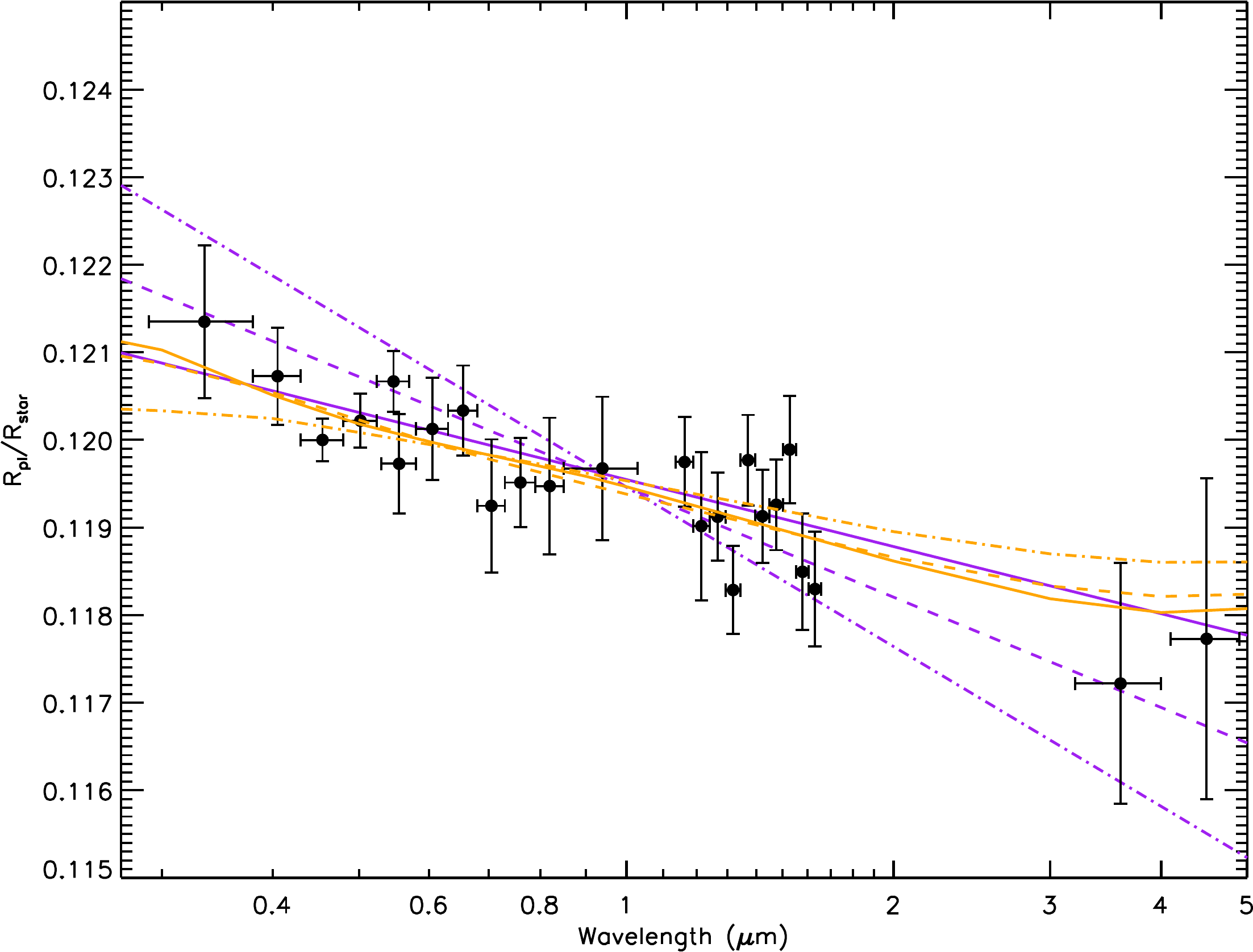}}
\caption[]{Rayleigh scattering (purple) and Mie scattering Al$_2$O$_3$
  models (orange) plotted at different assumed atmospheric temperatures of 968 K
  (solid), 1450 K (dashed), and 2100 K (dot-dashed).  Al$_2$O$_3$ Mie scattering
  models can provide good fits at a wide variety of temperature ranges, while
  Rayleigh scattering models fit best at temperatures of 882$\pm$164 K.}
\label{Figure:Dustfigure}
\end{figure}

As indicated in Table \ref{Table:fits}, our best fitting models are
ones which include significant opacity from a Rayleigh or Mie
scattering source.  Most of the expected aerosol-free models are ruled out
at high significance, limiting molecular absorption to a minor role in our broadband transmission spectrum.   
Potentially, Rayleigh scattering could be produced
by gaseous H$_2$, a possibility explored by \cite{2008A&A...485..865L} for HD~209458b and \cite{2013Icar..225..432S} 
when modelling the WASP-12b WFC3 emission and transmission spectra, though
there are several problems with this potential interpretation.  For
one, a true planetary atmosphere is unlikely to be devoid of all
atomic and molecular species, and there is evidence for substantial opacity
from metals in the near-UV \citep{2010ApJ...714L.222F}.  
Furthermore, a pure H$_2$ atmosphere would have a very high albedo,
which is ruled out by the {\it Spitzer} observations \citep{2012ApJ...747...82C}. 
Sub-micron silicate condensates could also be a source of Rayleigh
scattering, as observed for the haze in HD~189733b 
\citep{2013MNRAS.432.2917P}. 
However, the main problem with H$_2$ Rayleigh slope interpretations on
WASP-12b are the
low implied temperatures (882$\pm$165 K) which may not be feasible on
this very hot planet, measured to have brightness temperatures near
3000 K.  In contrast, the terminator temperatures implied by
the Rayleigh scattering slope for HD~189733b 
are perfectly consistent with day-side and phase curve measurements \citep{2008A&A...481L..83L, 2012ApJ...754...22K}.

It is possible the condensate scale height is smaller than that of the
surrounding gas, which would flatten the resulting transmission spectra
and could explain a low measured temperature for aerosols in the
Rayleigh scattering regime.
Solar system observations and brown dwarf models indicate condensate
scale heights which are $\sim$1/3 that of the local gas scale height
\citep{1994JGR....9914623C, 2002ApJ...568..335M}, a possibility for
hot-Jupiter atmospheres as well \citep{2005MNRAS.364..649F}.  
For WASP-12b, a Rayleigh scattering condensate with a scale height
1/3$H$ would imply temperatures 3$\times$ higher than found otherwise, or 2646$\pm$165 K
which is much closer to the expected terminator temperatures.
However, we note that the Rayleigh scattering signature on HD~189733b
is fully consistent with a condensate scale height which is the same as
the gas scale height \citep{2008A&A...481L..83L}. 

Our best fit Mie scattering aerosol models are indistinguishable in
quality to Rayleigh models, providing an excellent overall fit (see Table
\ref{Table:fits}).  Unlike the Rayleigh model, hotter temperature Mie scattering models can fit the
data equally well, which is likely necessary for WASP-12b (see Fig. \ref{Figure:Dustfigure}). 
Even with near-UV to IR wavelength coverage,
the particle size-temperature degeneracy makes it difficult to confidently
constrain the composition, since other aerosol materials with similar
optical properties to corundum would also provide equally good fits.  Despite
the strong degeneracy, we do find the particle sizes for all models
are generally sub-micron in size regardless of temperature.  

In principle, an aerosol could be
the result of 
condensation chemistry, or alternatively the aerosol may be photochemical in nature
\citep{2013arXiv1301.5627M}.  
If the aerosol observed arrises from condensation chemistry, WASP-12b would have
similarities with late M-dwarfs and brown dwarfs, both of which are
thought to have significant dust in their photospheres 
\citep{1997ApJ...480L..39J, 2006asup.book....1L,
  2008A&A...485..547H, 2013arXiv1301.5627M}. 

Hydrocarbons in hot Jupiters models have been explored in several
studies including \cite{2004ApJ...605L..61L},
\cite{2009arXiv0911.0728Z} and \cite{2011ApJ...737...15M}.  
These studies find that polycyclic aromatic hydrocarbon formation generally
occurs at high altitudes on the `cooler' hot-Jupiters, which could
ultimately lead to the formation of a `soot' aerosol layer \citep{2009arXiv0911.0728Z}.
Such a soot layer would be difficult to produce in current models
for WASP-12b, without invoking significant transport from the cooler
night side or pole regions.
In a model of HD~209458b, \cite{2004ApJ...605L..61L} 
found that C$_{\rm x}$H$_{\rm y}$ compounds were lost 
either primarily by reactions with atomic hydrogen or by
photolysis, and such process would likely be more
efficient in the higher irradiated conditions of WASP-12b.
However, as discussed in 
\cite{2011ApJ...737...15M}, 
the chemical models currently invoked assume neutral chemistry, which is not particularly conducive for soot formation,
even under `cooler' HD~189733b conditions, and it is possible that
ion chemistry plays a role in the formation of complex species, an
area which deserves further study.

There are other indications of significant aerosols, namely from
secondary eclipse and phase curve observations.  The emission spectrum
of WASP-12b resembles a black body from the optical to the IR
\citep{2012ApJ...760..140C}.  
Such a black body spectrum could be produced from an isothermal
atmosphere, but could also be explained by clouds.  For non-isothermal atmospheres, more consistent with
radiative equilibrium, an observed black body emission spectrum could naturally occur as a
result of a cloud deck.  The atmosphere could be optically thin at all
wavelengths at low pressure above the cloud deck, but then at the
pressure level of the cloud, the atmosphere may suddenly become optically thick, making
the surface emit from the same altitudes and temperatures
at all wavelengths.

The {\it Spitzer} phase curve measurements from  
\cite{2012ApJ...747...82C} 
also indicate a low heat recirculation efficiency and a modest Bond albedo,
estimated to be $A_B$=0.25$_{-0.10}^{+0.18}$.  As many hot Jupiters have very low
albedos \citep{2011ApJ...729...54C}, 
the modest albedo of WASP-12b
prompted \cite{2012ApJ...747...82C} 
to mention Rayleigh scattering and/or reflective clouds as a possibility.
Though the errors on albedo are currently large, if confirmed an aerosol species like
corundum could naturally produce such a modest albedo, as its optical
properties make the material not overly scattering (like MgSiO$_3$
or CaTiO$_3$)
nor highly absorptive (like Fe$_2$O$_3$).  Optical secondary eclipse
measurements could play an important role in determining the overall
pervasiveness of the aerosols, with a direct measurement of
the wavelength-dependant albedo a further important constraint on the
composition (e.g. see \citealt{2013arXiv1307.3239E} for the case of HD189733b).
For instance, a blue and red optical albedo measurement could help
differentiate between a strong Rayleigh scattering dust and tholins, which
are preferentially red.

Aerosol species could also be useful in interpreting the observed low recirculation
efficiency.  As pointed out by \cite{2012MNRAS.420...20H} and
\cite{2013MNRAS.432.2917P}, 
the first-order effect of clouds in hot Jupiters is to move the contribution functions
and effective photosphere at all wavelengths to higher altitudes and lower pressures.
At lower pressures, the timescale of heat loss from the
atmosphere will decrease, thereby also decreasing the efficiency of the observed day-to-night heat redistribution.


\subsection{The lack of TiO in very hot Jupiters}
To date, there has been no substantial evidence for TiO present in the
atmosphere for any hot Jupiter, except for the possible bump in the
absorption spectrum of HD~209458b between 6200A and $\sim$7000 \AA\
\citep{2008A&A...492..585D}.  
{\it HST} transmission spectra now shows
that TiO is absent in the atmospheres of the very hot Jupiters
WASP-12b and WASP-19b \citep{2013arXiv1307.2083H}.  
Both planets also lack strong thermal inversions
\citep{2012ApJ...760..140C,2013MNRAS.430.3422A},   
which further indicates TiO is not present in either planet at the necessary
  abundance levels to cause a stratosphere.  
There are four leading mechanisms to deplete TiO from
a hot Jupiter atmosphere: a vertical-cold trap, a day/night cold trap, the
effects of stellar activity and high C/O ratios \citep{2009ApJ...699.1487S, 2009ApJ...699..564S, 2010ApJ...720.1569K,  2012ApJ...758...36M}. 
As WASP-19b orbits an active star, experiences a strong day/night
contrast, and can also experience a vertical cold trap, it is not immediately clear
what the dominant TiO-depletion mechanism is for that planet, 
though a detection of H$_2$O in the transmission spectrum of WASP-19b disfavours the high C/O scenario \citep{2013arXiv1307.2083H}.  

WASP-12b is an ideal test case in many respects.
WASP-12A is inactive (see section \ref{section:activity}), making the activity scenario
unlikely.
Further, while most hot Jupiters can experience a vertical
cold-trap, which would deplete TiO at depth,  
WASP-12b is hot enough such that it should
not experience a cold-trap anywhere throughout its dayside
atmosphere \citep{2009ApJ...699.1487S}. 
Finally, sensitive non-detections of both TiO and TiH points towards a
  general lack of Ti-bearing molecules in the broadband spectrum, irrespective of the C/O ratio.

Very hot Jupiters, with equilibrium temperatures greater than 2400 K, are observed to have low circulation efficiencies
which leads to very large day/night temperature contrasts
\citep{2012ApJ...747...82C}.   
The efficiency of redistribution is primarily governed by the ratio of
advective to radiative timescales, with hotter
planets tend to have shorter
radiative timescales and thus larger day/night contrasts
\citep{2002A&A...385..166S,2012ApJ...751...59P, 2012ApJ...747...82C, 2013arXiv1306.4673P}. 
Night-side temperatures cool enough for species such as Ti to be condensed into
solids could lead to rain out, which would then drastically deplete
the species from the atmosphere.
\cite{2013arXiv1301.4522P} 
has explored how a strong day/night temperature contrast can lead to
depletion of TiO through the use of 3D modelling.  In their models,
they found that large-scale circulation patterns can produce strong vertical mixing that can keep
condensable species lofted, as long as they are trapped in particles of
sizes of a few microns or less on the night side.  Of note, our transmission
spectrum is consistent with sub-micron size particles.  

Thus, while it is not entirely clear if the aerosols at the terminator
originate from condensation chemistry, in any case, the day/night cold-trap would appear to be the leading
explanation for the lack of TiO.


\section{CONCLUSION}

WASP-12b is now among only a small handful of exoplanets with a full
optical to near-IR transmission spectrum measured.  
With high S/N broadband observations reaching into the near-UV, we are
able to decisively rule out 
prominent absorption by TiO
in the exoplanet's atmosphere.  Given WASP-12b's unique
properties, we find that the day/night cold trap is
the leading explanation for the lack of the molecule.  As other very
hot Jupiters also experience similarly strong day/night contrasts, it is
likely that few (if any) hot Jupiters will show strong TiO absorption features.

We find a transmission spectrum which lacks any strong broadband molecular or atomic
absorption signatures, though it does have a significant slope indicative of
scattering by aerosols.  This is the first direct indication for the
presence of aerosols on a very hot Jupiter.  In addition to the transmission spectrum, significant aerosols throughout the entire
atmosphere can also help explain the black body nature of the emission
spectrum, as well as the modestly bright albedo.  An aerosol species on WASP-12b could originate from a
high temperature condense analogous to the dusts on brown dwarfs and
M-dwarfs, or alternatively the aerosols could be hydrocarbons which are photochemical in nature.

To date, only the much cooler planet HD~189733b has strong
evidence for aerosol species in the atmosphere of a hot Jupiter.
The results here for the much hotter planet WASP-12b indicate that the aerosol
species can potentially be an important feature
for planets across the entire wide range of hot Jupiter planets.

Transmission spectra of exoplanets provides a unique and powerful
measurement, with the capability to help identify and constrain
high temperature dust species or photochemical products.  In
this regard, transmission spectra have the potential to also help our understanding of
brown dwarfs and M-dwarfs, of which such measurements are not possible.
In the future, the mid infrared capabilities of MIRI on JWST could
detect the major dust absorption features, which could directly
identify aerosol composition.

For the near future, optical albedo measurements can independently
help confirm the presence of a reflecting cloud and constrain its composition.  In addition, a measure (or model constraints) of the terminator temperature can
help constrain the particle composition, and additional thermal phase
curve information will be helpful in this regard.

\section*{ACKNOWLEDGMENTS}

This work is based on observations
with the NASA/ESA {\it Hubble Space Telescope}, obtained at the
Space Telescope Science Institute (STScI) operated by AURA, Inc.  
Support for this work was provided by NASA through grants under the HST-GO-12473 programme from
the STScI.  
We thank I. Baraffe for useful discussions, and the anonymous referee
for their comments.
We also warmly thank Jason W. Ferguson for providing the optical constants
for CaTiO$_3$.
C. Huitson, P. Wilson and H. Wakeford acknowledge support
from the UK Science \& Technology Facilities Council (STFC). 
D. Sing, F. Pont and, N. Nikolov acknowledge support from STFC consolidated grant
ST/J0016/1.
A. Lecavelier and A. Vidal-Madjar acknowledge support from the French Agence Nationale de la Recherche (ANR), under program ANR-12-BS05-0012 ``Exo-Atmos''.
\footnotesize{
\bibliographystyle{mn2e} 
\bibliography{wasp12.manuscript.astroph} 
}
\end{document}